\documentclass[11pt,twoside]{article}

\usepackage{graphics}
\usepackage{natbib}
\usepackage{amsmath}

\renewcommand{\sectionmark}[1]{\markboth{\textit{Random graphs as models
of networks}}{\textit{#1}}}
\renewcommand{\subsectionmark}[1]{\markboth{\textit{Random graphs as models
of networks}}{\textit{#1}}}

\newcommand{\captionfonts}{\small}
\makeatletter
\long\def\@makecaption#1#2{%
  \vskip\abovecaptionskip
  \sbox\@tempboxa{{\captionfonts #1: #2}}%
  \ifdim \wd\@tempboxa >\hsize
    {\captionfonts #1: #2\par}
  \else
    \hbox to\hsize{\hfil\box\@tempboxa\hfil}%
  \fi
  \vskip\belowcaptionskip}
\makeatother

\newcommand{\defn}{\textbf}
\newcommand{\half}{\mbox{$\frac12$}}
\newcommand{\citey}{\citeyearpar}
\renewcommand{\d}{{\rm d}}
\renewcommand{\i}{{\rm i}}
\renewcommand{\O}{{\rm O}}
\newcommand{\e}{{\rm e}}
\newcommand{\set}[1]{\lbrace#1\rbrace}
\newcommand{\av}[1]{\langle#1\rangle}
\newcommand{\eref}[1]{(\ref{#1})}
\newcommand{\etal}{{\it{}et~al.}}
\newcommand{\cG}{\mathcal{G}}

\begin{document}

\title{Random graphs as models of networks}
\author{M. E. J. Newman\\
\textit{\small Santa Fe Institute, 1399 Hyde Park Road, Santa Fe,
NM 87501, U.S.A.}}
\date{}
\maketitle

\begin{abstract}
  The random graph of Erd\H{o}s and R\'enyi is one of the oldest and best
  studied models of a network, and possesses the considerable advantage of
  being exactly solvable for many of its average properties.  However, as a
  model of real-world networks such as the Internet, social networks or
  biological networks it leaves a lot to be desired.  In particular, it
  differs from real networks in two crucial ways: it lacks network
  clustering or transitivity, and it has an unrealistic Poissonian degree
  distribution.  In this paper we review some recent work on
  generalizations of the random graph aimed at correcting these
  shortcomings.  We describe generalized random graph models of both
  directed and undirected networks that incorporate arbitrary non-Poisson
  degree distributions, and extensions of these models that incorporate
  clustering too.  We also describe two recent applications of random graph
  models to the problems of network robustness and of epidemics spreading
  on contact networks.
\end{abstract}

\section{Introduction}
\thispagestyle{empty}
In a series of seminal papers in the 1950s and 1960s, Paul Erd\H{o}s and
Alfr\'ed R\'enyi proposed and studied one of the earliest theoretical
models of a network, the \defn{random graph} \citep{ER59,ER60,ER61}.  This
minimal model consists of $n$ nodes or \defn{vertices}, joined by links or
\defn{edges} which are placed between pairs of vertices chosen uniformly at
random.  Erd\H{o}s and R\'enyi gave a number of versions of their model.
The most commonly studied is the one denoted $G_{n,p}$, in which each
possible edge between two vertices is present with independent
probability~$p$, and absent with probability~$1-p$.  Technically, in fact,
$G_{n,p}$ is the \emph{ensemble} of graphs of $n$ vertices in which each
graph appears with the probability appropriate to its number of
edges.\footnote{For a graph with $n$ vertices and $m$ edges this
  probability is $p^m(1-p)^{M-m}$, where $M=\half n(n-1)$.}

Often one wishes to express properties of $G_{n,p}$ not in terms of $p$ but
in terms of the average degree $z$ of a vertex.  (The \defn{degree} of a
vertex is the number of edges connected to that vertex.)  The average
number of edges on the graph as a whole is $\half n(n-1)p$, and the average
number of \emph{ends} of edges is twice this, since each edge has two ends.
So the average degree of a vertex is
\begin{equation}
z = {n(n-1)p\over n} = (n-1)p \simeq np,
\end{equation}
where the last approximate equality is good for large~$n$.  Thus, once we
know~$n$, any property that can be expressed in terms of~$p$ can also be
expressed in terms of~$z$.

The Erd\H{o}s--R\'enyi random graph has a number of desirable properties as
a model of a network.  In particular it is found that many of its ensemble
average properties can be calculated exactly in the limit of large~$n$
\citep{Bollobas85,JLR99}.  For example, one interesting feature, which was
demonstrated in the original papers by Erd\H{o}s and R\'enyi, is that the
model shows a phase transition\footnote{Erd\H{o}s and R\'enyi didn't call
  it that, but that's what it is.} with increasing $z$ at which a
\defn{giant component} forms.  A \defn{component} is a subset of vertices
in the graph each of which is reachable from the others by some path
through the network.  For small values of $z$, when there are few edges in
the graph, it is not surprising to find that most vertices are disconnected
from one another, and components are small, having an average size that
remains constant as the graph becomes large.  However, there is a critical
value of $z$ above which the one largest component in the graph contains a
finite fraction $S$ of the total number of vertices, i.e.,~its size $nS$
scales linearly with the size of the whole graph.  This largest component
is the giant component.  In general there will be other components in
addition to the giant component, but these are still small, having an
average size that remains constant as the graph grows larger.  The phase
transition at which the giant component forms occurs precisely at $z=1$.
If we regard the fraction $S$ of the graph occupied by the largest
component as an order parameter, then the transition falls in the same
universality class as the mean-field percolation transition \citep{SA92}.

The formation of a giant component in the random graph is reminiscent of
the behaviour of many real-world networks.  One can imagine loose-knit
networks for which there are so few edges that, presumably, the network has
no giant component, and all vertices are connected to only a few others.
The social network in which pairs of people are connected if they have had
a conversation within the last 60 seconds, for example, is probably so
sparse that it has no giant component.  The network in which people are
connected if they have \emph{ever} had a conversation, on the other hand,
is very densely connected and certainly has a giant component.

However, the random graph differs from real-world networks in some
fundamental ways also.  Two differences in particular have been noted in
the recent literature \citep{Strogatz01,AB02}.  First, as pointed out by
Watts and Strogatz~(\citeyear{WS98}; \citealt{Watts99a}) real-world
networks show strong \defn{clustering} or \defn{network transitivity},
where Erd\H{o}s and R\'enyi's model does not.  A network is said to show
clustering if the probability of two vertices being connected by an edge is
higher when the vertices in question have a common neighbour.  That is,
there is another vertex in the network to which they are both attached.
Watts and Strogatz measured this clustering by defining a \defn{clustering
  coefficient}~$C$, which is the average probability that two neighbours of
a given vertex are also neighbours of one another.  In many real-world
networks the clustering coefficient is found to have a high value, anywhere
from a few percent to 50 percent or even more.  In the random graph of
Erd\H{o}s and R\'enyi on the other hand, the probabilities of vertex pairs
being connected by edges are by definition independent, so that there is no
greater probability of two vertices being connected if they have a mutual
neighbour than if they do not.  This means that the clustering coefficient
for a random graph is simply $C=p$, or equivalently $C\simeq z/n$.  In
Table~\ref{clustering} we compare clustering coefficients for a number of
real-world networks with their values on a random graph with the same
number of vertices and edges.  The graphs listed in the table are:
\begin{table}[t]
\begin{center}
\begin{tabular}{l|rr|ll}
 & & & \multicolumn{2}{c}{clustering coefficient $C$} \\
 network                                 &      $n$      & $z$ & measured  & random graph \\
\hline
 Internet (autonomous systems)$^{\rm a}$ & $6\,374$      & $3.8$          & $0.24$         & $0.00060$    \\
 World-Wide Web (sites)$^{\rm b}$        & $153\,127$    & $35.2$         & $0.11$    & $0.00023$    \\
 power grid$^{\rm c}$                    & $4\,941$      & $2.7$          & $0.080$   & $0.00054$    \\
 biology collaborations$^{\rm d}$        & $1\,520\,251$ & $15.5$         & $0.081$   & $0.000010$   \\
 mathematics collaborations$^{\rm e}$    & $253\,339$    & $3.9$          & $0.15$    & $0.000015$   \\
 film actor collaborations$^{\rm f}$     & $449\,913$    & $113.4$        & $0.20$    & $0.00025$    \\
 company directors$^{\rm f}$             & $7\,673$      & $14.4$         & $0.59$    & $0.0019$     \\
 word co-occurrence$^{\rm g}$            & $460\,902$    & $70.1$         & $0.44$    & $0.00015$    \\
 neural network$^{\rm c}$                & $282$         & $14.0$         & $0.28$    & $0.049$      \\
 metabolic network$^{\rm h}$             & $315$         & $28.3$         & $0.59$    & $0.090$      \\
 food web$^{\rm i}$                      & $134$         & $8.7$          & $0.22$    & $0.065$
\end{tabular}
\end{center}
\caption{Number of vertices~$n$, mean degree~$z$, and clustering
coefficient~$C$ for a number of different networks.  Numbers are taken
from $^{\rm a}$\citet{PVV01}, $^{\rm b}$\citet{Adamic99}, 
$^{\rm c}$\citet{WS98}, $^{\rm d}$\citet{Newman01b}, 
$^{\rm e}$\citet{Newman02a}, $^{\rm f}$\cite{NSW01},
$^{\rm g}$\citet{CS01}, $^{\rm h}$\citet{FW00}, $^{\rm i}$\citet{MS02}.}
\label{clustering}
\end{table}
\begin{itemize}
\setlength{\itemsep}{0pt}
\item Internet: a graph of the fibre optic connections that comprise the
  Internet, at the level of so-called ``autonomous systems.''  An
  autonomous system is a group of computers within which data flow is
  handled autonomously, while data flow between groups is conveyed over the
  public Internet.  Examples of autonomous systems might be the computers
  at a company, a university, or an Internet service provider.
\item World-Wide Web: a graph of sites on the World-Wide Web in which edges
  represent ``hyperlinks'' connecting one site to another.  A site in this
  case means a collection of pages residing on a server with a given name.
  Although hyperlinks are directional, their direction has been ignored in
  this calculation of the clustering coefficient.
\item Power grid: a graph of the Western States electricity transmission
  grid in the United States.  Vertices represent stations and substations;
  edges represent transmission lines.
\item Biology collaborations: a graph of collaborations between researchers
  working in biology and medicine.  A collaboration between two scientists
  is defined in this case as coauthorship of a paper that was catalogued in
  the Medline bibliographic database between 1995 and 1999 inclusive.
\item Mathematics collaborations: a similar collaboration graph for
  mathematicians, derived from the archives of \emph{Mathematical
    Reviews}.
\item Film actor collaborations: a graph of collaborations between film
  actors, where a collaboration means that the two actors in question have
  appeared in a film together.  The data are from the Internet Movie
  Database.
\item Company directors: a collaboration graph of the directors of
  companies in the Fortune 1000 for 1999.  (The Fortune 1000 is the 1000 US
  companies with the highest revenues during the year in question.)
  Collaboration in this case means that two directors served on the board
  of a Fortune 1000 company together.
\item Word co-occurrences: a graph in which the vertices represent words in
  the English language, and an edge signifies that the vertices it connects
  frequently occur in adjacent positions in sentences.
\item Neural network: a graph of the neural network of the worm
    \emph{C.\ Elegans.}
\item Metabolic network: a graph of interactions forming a part of the
  energy generation and small building block synthesis metabolism of the
  bacterium \emph{E. Coli.}  Vertices represent substrates and products,
  and edges represent interactions.
\item Food web: the food web of predator--prey interactions between species
  in Ythan Estuary, a marine estuary near Aberdeen, Scotland.  Like the
  links in the World-Wide Web graph, the directed nature of the
  interactions in this food web have been neglected for the purposes of
  calculating the clustering coefficient.
\end{itemize}

As the table shows, the agreement between the clustering coefficients in
the real networks and in the corresponding random graphs is not good.  The
real and theoretical figures differ by as much as four orders of magnitude
in some cases.  Clearly, the random graph does a poor job of capturing this
particular property of networks.

\begin{figure}
\hbox to\textwidth{\resizebox{6.8cm}{!}{\includegraphics{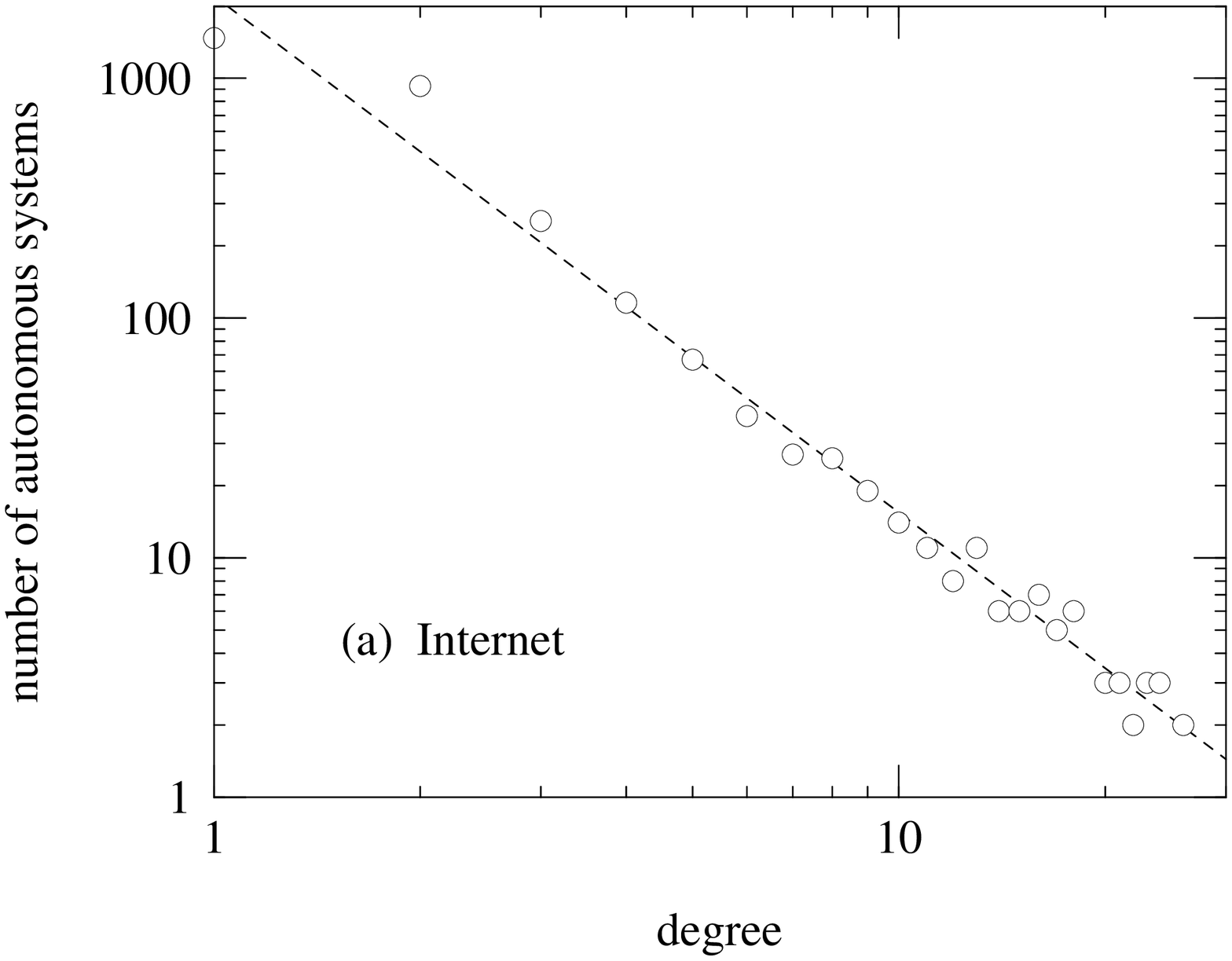}}\hfil
                   \resizebox{6.8cm}{!}{\includegraphics{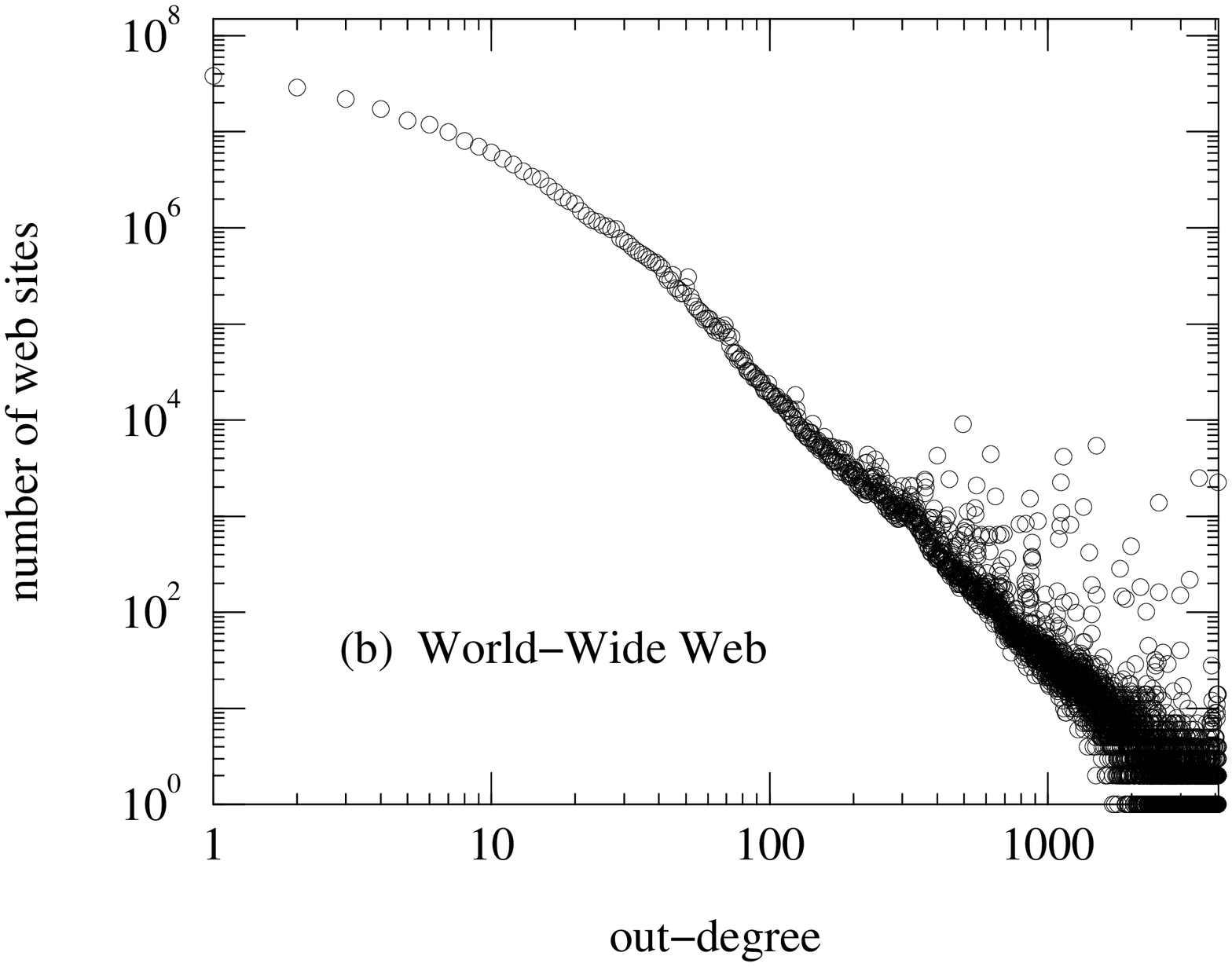}}}
\vspace{5mm}
\hbox to\textwidth{\resizebox{6.8cm}{!}{\includegraphics{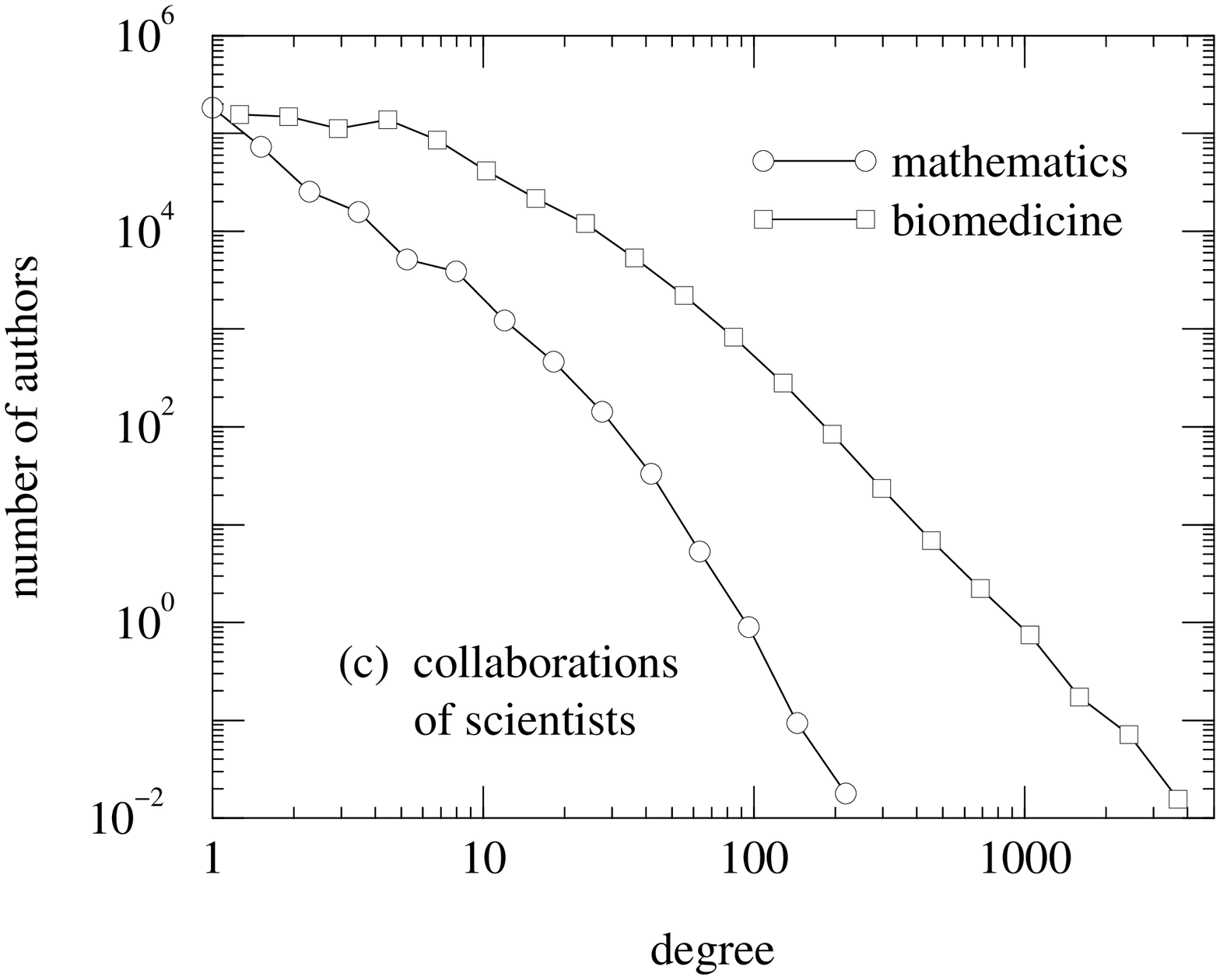}}\hfil
                   \resizebox{6.8cm}{!}{\includegraphics{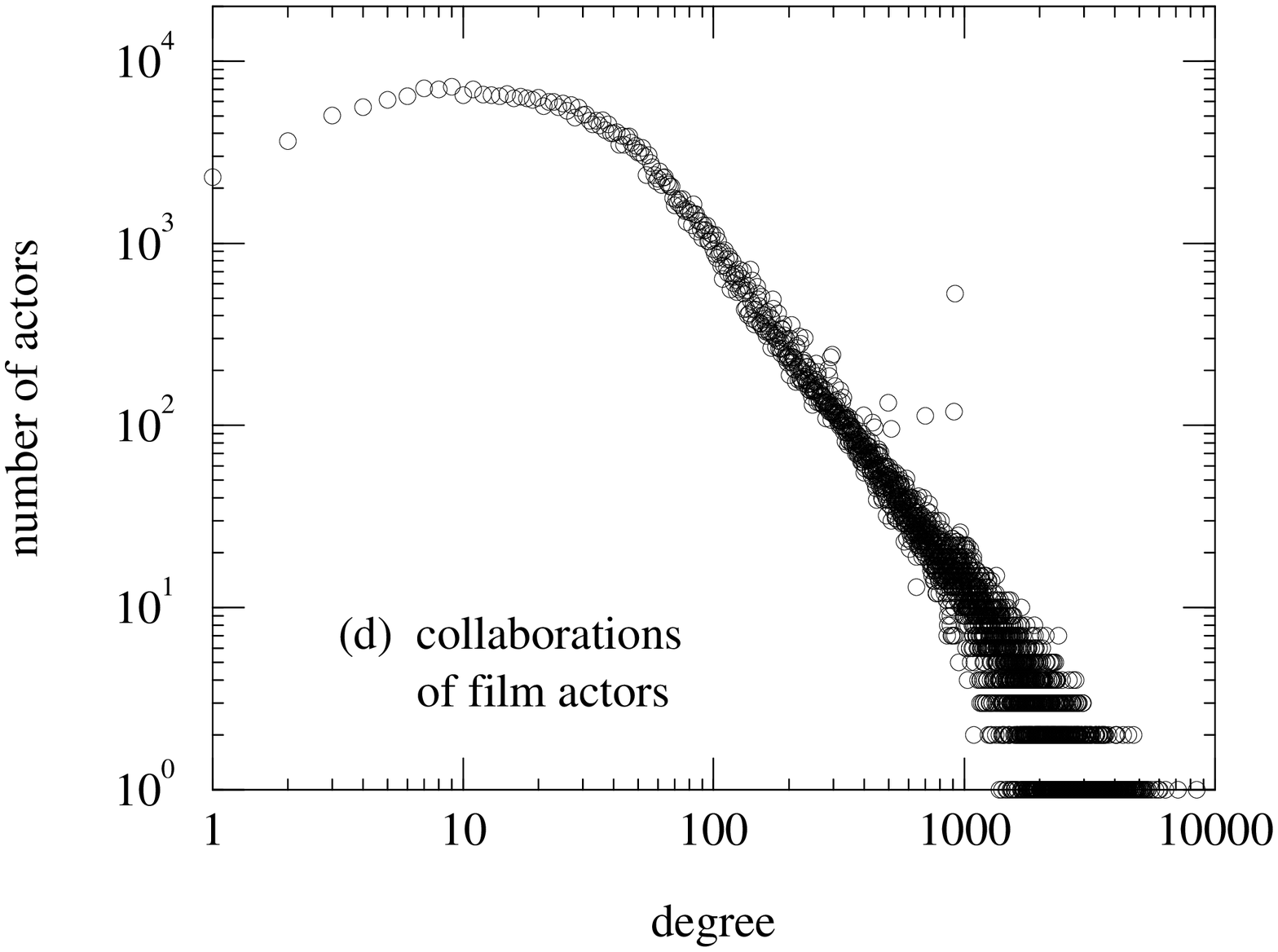}}}
\vspace{5mm}
\hbox to\textwidth{\resizebox{6.8cm}{!}{\includegraphics{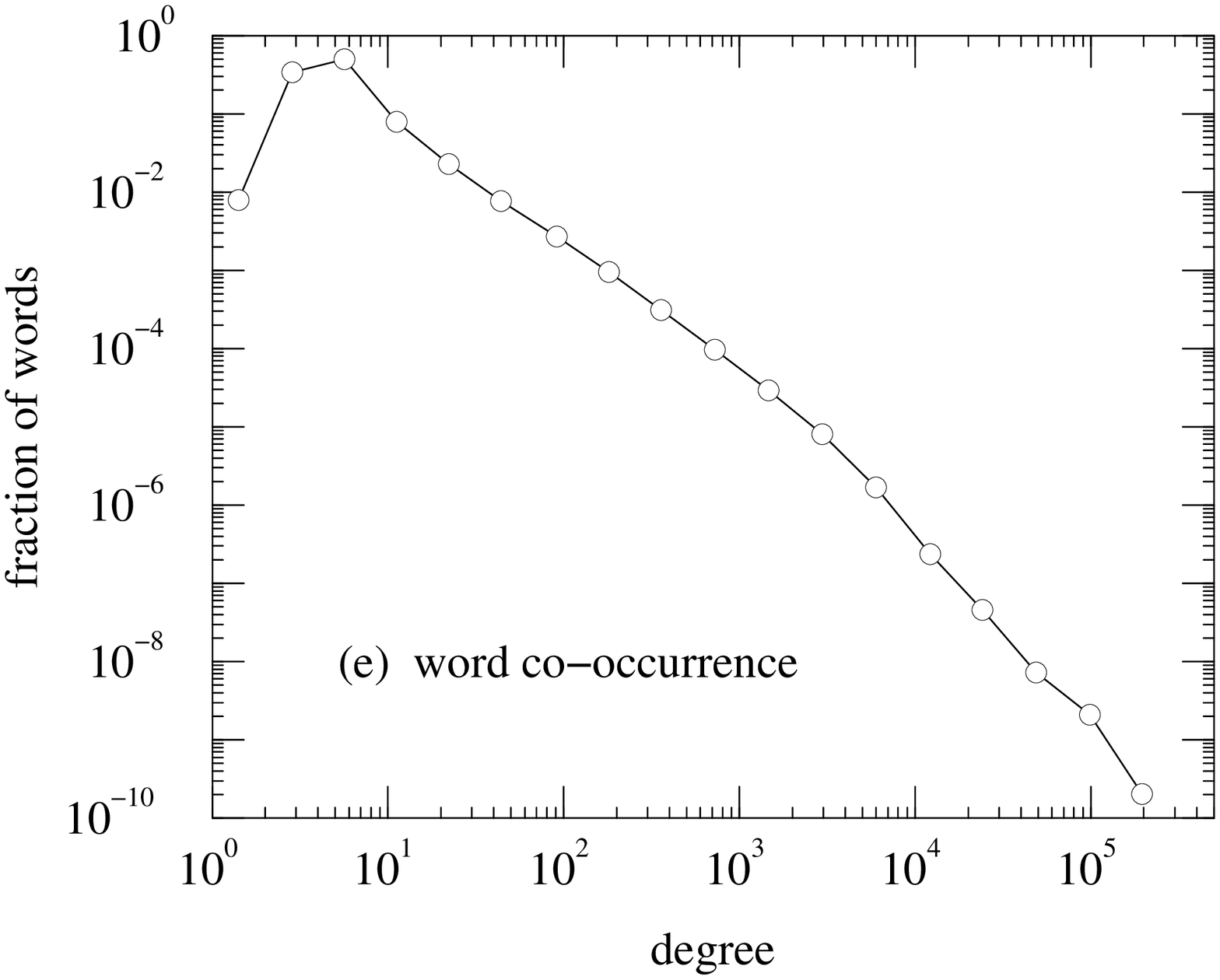}}\hfil
                   \resizebox{6.8cm}{!}{\includegraphics{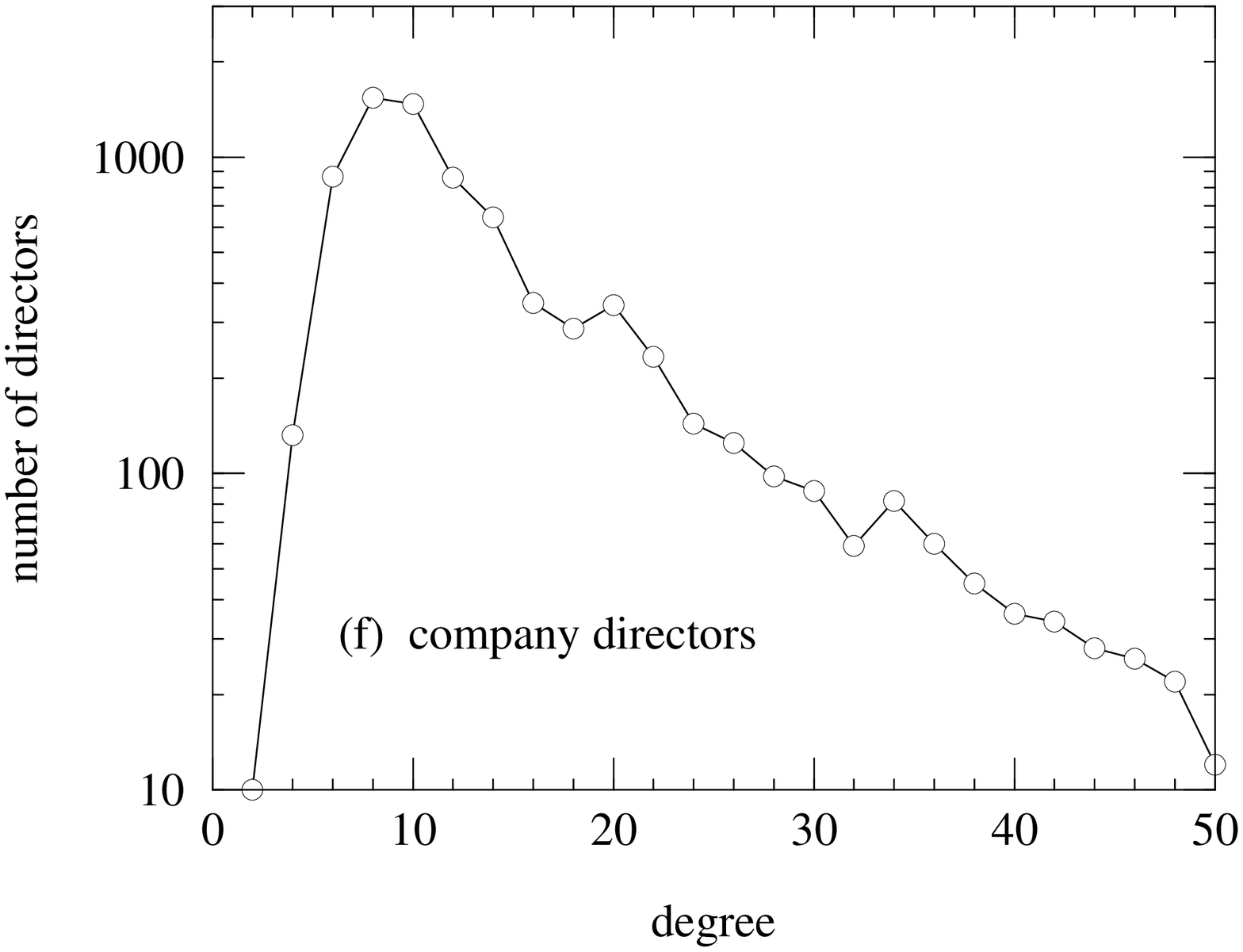}}}
\caption{Measured degree distributions for a number of different networks.
  (a)~Physical connections between autonomous systems on the Internet,
  \textit{circa} 1997 \citep{FFF99}.  (b)~A 200 million page subset of the
  World-Wide Web, \textit{circa} 1999 \citep{Broder00}.  The figure shows
  the out-degree of pages, i.e.,~numbers of links pointing from those pages
  to other pages.  (c)~Collaborations between biomedical scientists and
  between mathematicians \citep{Newman01b,Newman02a}.  (d)~Collaborations
  of film actors \citep{ASBS00}.  (e)~Co-occurrence of words in the English
  language \citep{CS01}.  (f)~Board membership of directors of Fortune 1000
  companies for year 1999 \citep{NSW01}.}
\label{degree}
\end{figure}

A second way in which random graphs differ from their real-world
counterparts is in their degree distributions, a point which has been
emphasized particularly in the work of Albert, Barab\'asi, and
collaborators \citep{AJB99,BA99b}.  The probability $p_k$ that a
vertex in an Erd\H{o}s--R\'enyi random graph has degree exactly~$k$ is
given by the binomial distribution:
\begin{equation}
p_k = {n-1\choose k} p^k (1-p)^{n-1-k}.
\end{equation}
In the limit where $n\gg kz$, this becomes
\begin{equation}
p_k = {z^k \e^{-z}\over k!},
\label{poisson}
\end{equation}
which is the well-known Poisson distribution.  Both binomial and Poisson
distributions are strongly peaked about the mean~$z$, and have a large-$k$
tail that decays rapidly as $1/k!$.  We can compare these predictions to
the degree distributions of real networks by constructing histograms of the
degrees of vertices in the real networks.  We show some examples, taken
from the networks described above, in Fig.~\ref{degree}.  As the figure
shows, in most cases the degree distribution of the real network is very
different from the Poisson distribution.  Many of the networks, including
Internet and World-Wide Web graphs, appear to have power-law degree
distributions \citep{AJB99,FFF99,Broder00}, which means that a small but
non-negligible fraction of the vertices in these networks have very large
degree.  This behaviour is quite unlike the rapidly decaying Poisson degree
distribution, and can have profound effects on the behaviour of the
network, as we will see later in this paper.  Other networks, particularly
the collaboration graphs, appear to have power-law degree distributions
with an exponential cutoff at high degree
\citep{ASBS00,Newman01a,Newman01b}, while others still, such as the graph
of company directors, seem to have degree distributions with a purely
exponential tail \citep{NSW01}.  The power grid of Table~\ref{clustering}
is another example of a network that has an exponential degree distribution
\citep{ASBS00}.

In this paper we show how to generalize the Erd\H{o}s--R\'enyi random graph
to mimic the clustering and degree properties of real-world networks.  In
fact, most of the paper is devoted to extensions that correct the degree
distribution, for which an elegant body of theory has been developed in the
last few years.  However, towards the end of the paper we also consider
ways in which clustering can be introduced into random graphs.  Work on
this latter problem is significantly less far advanced than work on degree
distributions, and we have at present only a few preliminary results.
Whether these results can be extended, and how, are open questions.

\section{Random graphs with specified degree distributions}
\label{rg}
It is relatively straightforward to generate random graphs that have
non-Poisson degree distributions.  The method for doing this has been
discussed a number of times in the literature, but appears to have been put
forward first by Bender and Canfield~\citey{BC78}.  The trick is to
restrict oneself to a specific degree \emph{sequence}, i.e.,~to a specified
set $\set{k_i}$ of the degrees of the vertices $i=1\ldots n$.  Typically
this set will be chosen in such a way that the fraction of vertices having
degree $k$ will tend to the desired degree distribution $p_k$ as $n$
becomes large.  For practical purposes however, such as numerical
simulation, it is almost always adequate simply to draw a degree sequence
$\set{k_i}$ from the distribution $p_k$ directly.

Once one has one's degree sequence, the method for generating the graph is
as follows: one gives each vertex $i$ a number $k_i$ of ``stubs''---ends of
edges emerging from the vertex---and then one chooses pairs of these stubs
uniformly at random and joins them together to make complete edges.  When
all stubs have been used up, the resulting graph is a random member of the
ensemble of graphs with the desired degree sequence.\footnote{The only
  small catch to this algorithm is that the total number of stubs must be
  even if we are not to have one stub left over at the end of the pairing
  process.  Thus we should restrict ourselves to degree sequences for which
  $\sum_i k_i$ is even.}  Note that, because of the $k_i!$ possible
permutations of the stubs emerging from the $i$th vertex, there are
$\prod_i k_i!$ different ways of generating each graph in the ensemble.
However, this factor is constant so long as the degree sequence $\set{k_i}$
is held fixed, so it does not prevent the method from sampling the ensemble
correctly.  This is the reason why we restrict ourselves to a fixed degree
sequence---merely fixing the degree distribution is not adequate to ensure
that the method described here generates graphs uniformly at random from
the desired ensemble.

The method of Bender and Canfield does not allow us to specify a clustering
coefficient for our graph.  (The clustering coefficient had not been
invented yet when Bender and Canfield were writing in 1978.)  Indeed the
fact that the clustering coefficient is not specified is one of the crucial
properties of these graphs that makes it possible, as we will show, to
solve exactly for many of their properties in the limit of large graph
size.  As an example of why this is important, consider the following
simple calculation.  The mean number of neighbours of a randomly chosen
vertex~A in a graph with degree distribution~$p_k$ is $z=\av{k}=\sum_k
kp_k$.  Suppose however that we want to know the mean number of second
neighbours of vertex~A, i.e.,~the mean number of vertices two steps away
from~A in the graph.  In a network with clustering, many of the second
neighbours of a vertex are also first neighbours---the friend of my friend
is also my friend---and we would have to allow for this effect to order
avoid overcounting the number of second neighbours.  In our random graphs
however, no allowances need be made.  The probability that one of the
second neighbours of~A is also a first neighbour goes as $n^{-1}$ in the
random graph, regardless of degree distribution, and hence can be ignored
in the limit of large~$n$.

There is another effect, however, that we certainly must take into account
if we wish to compute correctly the number of second neighbours: the degree
distribution of the first neighbour of a vertex is not the same as the
degree distribution of vertices on the graph as a whole.  Because a
high-degree vertex has more edges connected to it, there is a higher chance
that any given edge on the graph will be connected to it, in precise
proportion to the vertex's degree.  Thus the probability distribution of
the degree of the vertex to which an edge leads is proportional to $kp_k$
and not just~$p_k$ \citep{Feld91,MR95,Newman02a}.  This distinction is
absolutely crucial to all the further developments of this paper, and the
reader will find it worthwhile to make sure that he or she is comfortable
with it before continuing.

In fact, we are interested here not in the complete degree of the vertex
reached by following an edge from~A, but in the number of edges emerging
from such a vertex other than the one we arrived along, since the latter
edge only leads back to vertex~A and so does not contribute to the number
of second neighbours of~A.  This number is one less than the total degree
of the vertex and its correctly normalized distribution is therefore
$q_{k-1}=kp_k/\sum_j jp_j$, or equivalently
\begin{equation}
q_k = {(k+1) p_{k+1}\over\sum_j j p_j}.
\label{defsqk}
\end{equation}
The average degree of such a vertex is then
\begin{equation}
\sum_{k=0}^\infty k q_k
  = {\sum_{k=0}^\infty k(k+1)p_{k+1}\over\sum_j j p_j}
  = {\sum_{k=0}^\infty (k-1)kp_k\over\sum_j j p_j}
  = {\av{k^2}-\av{k}\over\av{k}}.
\label{avqk}
\end{equation}
This is the average number of vertices two steps away from our vertex~A via
a particular one of its neighbours.  Multiplying this by the mean degree
of~A, which is just $z=\av{k}$, we thus find that the mean number of second
neighbours of a vertex is
\begin{equation}
z_2 = \av{k^2} - \av{k}.
\label{z2}
\end{equation}
If we evaluate this expression using the Poisson degree distribution,
Eq.~\eref{poisson}, then we get $z_2=\av{k}^2$---the mean number of second
neighbours of a vertex in an Erd\H{o}s--R\'enyi random graph is just the
square of the mean number of first neighbours.  This is a special case
however.  For most degree distributions Eq.~\eref{z2} will be dominated by
the term $\av{k^2}$, so the number of second neighbours is roughly the mean
square degree, rather than the square of the mean.  For broad distributions
such as those seen in Fig.~\ref{degree}, these two quantities can be very
different \citep{Newman02a}.

We can extend this calculation to further neighbours also.  The average
number of edges leading from each second neighbour, other than the one we
arrived along, is also given by~\eref{avqk}, and indeed this is true at any
distance $m$ away from vertex~A.  Thus the average number of neighbours at
distance $m$ is
\begin{equation}
z_m = {\av{k^2}-\av{k}\over\av{k}} z_{m-1} = {z_2\over z_1} z_{m-1},
\end{equation}
where $z_1\equiv z=\av{k}$ and $z_2$ is given by Eq.~\eref{z2}.  Iterating
this equation we then determine that
\begin{equation}
z_m = \biggl[{z_2\over z_1}\biggr]^{m-1} z_1.
\label{zm}
\end{equation}
Depending on whether $z_2$ is greater than $z_1$ or not, this expression
will either diverge or converge exponentially as $m$ becomes large, so that
the average total number of neighbours of vertex A at all distances is
finite if $z_2<z_1$ or infinite if $z_2>z_1$ (in the limit of
infinite~$n$).\footnote{The case of $z_1=z_2$ is deliberately missed out
  here, since it is non-trivial to show how the graph behaves exactly at
  this transition point \citep{Bollobas85}.  For our current practical
  purposes however, this matters little, since the chances of any real
  graph being precisely at the transition point are negligible.}  If this
number is finite, then clearly there can be no giant component in the
graph.  Conversely, if it is infinite, then there must be a giant
component.  Thus the graph shows a phase transition similar to that of the
Erd\H{o}s--R\'enyi graph precisely at the point where $z_2=z_1$.  Making
use of Eq.~\eref{z2} and rearranging, we find that this condition is also
equivalent to $\av{k^2}-2\av{k}=0$, or, as it is more commonly written,
\begin{equation}
\sum_{k=0}^\infty k(k-2) p_k = 0.
\label{mrcondition}
\end{equation}
This condition for the position of the phase transition in a random graph
with arbitrary degree sequence was first given by Molloy and
Reed~\citey{MR95}.

An interesting feature of Eq.~\eref{mrcondition} is that, because of the
factor $k(k-2)$, vertices of degree zero and degree two contribute nothing
to the sum, and therefore the number of such vertices does not affect the
position of the phase transition or the existence of the giant component.
It is easy to see why this should be the case for vertices of degree zero;
obviously one can remove (or add) degree-zero vertices without changing the
fact of whether a giant component does or does not exist in a graph.  But
why vertices of degree two?  This has a simple explanation also: removing
vertices of degree two does not change the topological structure of a graph
because all such vertices fall in the middle of edges between other pairs
of vertices.  We can therefore remove (or add) any number of such vertices
without affecting the existence of the giant component.

Another quantity of interest in many networks is the typical distance
through the network between pairs of vertices
\citep{Milgram67,TM69,PK78,WS98,ASBS00}.  We can use Eq.~\eref{zm} to make
a calculation of this quantity for our random graph as follows.  If we are
``below'' the phase transition of Eq.~\eref{mrcondition}, in the regime
where there is no giant component, then most pairs of vertices will not be
connected to one another at all, so vertex--vertex distance has little
meaning.  Well above the transition on the other hand, where there is a
giant component, all vertices in this giant component are connected by some
path to all others.  Eq.~\eref{zm} tells us the average number of vertices
a distance $m$ away from a given vertex~A in the giant component.  When the
total number of vertices within distance $m$ is equal to the size $n$ of
the whole graph, $m$~is equal to the so-called ``radius''~$r$ of the
network around vertex~A.  Indeed, since $z_2/z_1\gg1$ well above the
transition, the number of vertices at distance $m$ grows quickly with~$m$
in this regime (see Eq.~\eref{zm} again), which means that most of the
vertices in the network will be far from~A, around distance~$r$, and $r$ is
thus also approximately equal to the average vertex--vertex
distance~$\ell$.  Well above the transition therefore, $\ell$ is given
approximately by $z_\ell\simeq n$, or
\begin{equation}
\ell = {\log (n/z_1)\over\log (z_2/z_1)} + 1.
\label{ell}
\end{equation}
For the special case of the Erd\H{o}s--R\'enyi random graph, for which
$z_1=z$ and $z_2=z^2$ as noted above, this expression reduces to the
well-known standard formula for this case: $\ell=\log n/\log z$
\citep{Bollobas85}.

The important point to notice about Eq.~\eref{ell} is that the
vertex--vertex distance increases logarithmically with the graph size~$n$,
i.e.,~it grows rather slowly.\footnote{Krzywicki~\citey{Krzywicki01} points
  out that this is true only for components such as the giant component
  that contain loops.  For tree-like components that contain no loops the
  mean vertex--vertex distance typically scales as a power of~$n$.  Since
  the giant components of neither our models nor our real-world networks
  are tree-like, however, this is not a problem.}  Even for very large
networks we expect the typical distance through the network from one vertex
to another to be quite small.  In social networks this effect is known as
the \defn{small-world effect},\footnote{Some authors, notably Watts and
  Strogatz~\citey{WS98}, have used the expression ``small-world network''
  to refer to a network that simultaneously shows both the small-world
  effect and high clustering.  To prevent confusion however we will avoid
  this usage here.}  and was famously observed by the experimental
psychologist Stanley Milgram in the letter-passing experiments he conducted
in the 1960s \citep{Milgram67,TM69,Kleinfeld02}.  More recently it has been
observed also in many other networks including non-social networks
\citep{WS98,ASBS00}.  This should come as no great surprise to us however.
On the contrary, it would be surprising if most networks did \emph{not}
show the small-world effect.  If we define the \defn{diameter} $d$ of a
graph to be the \emph{maximum} distance between any two connected vertices
in the graph, then it can be proven rigorously that the fraction of all
possible graphs with $n$ vertices and $m$ edges for which $d>c\log n$ for
some constant $c$ tends to zero as $n$ becomes large \citep{Bollobas85}.
And clearly if the diameter increases as $\log n$ or slower, then so also
must the average vertex--vertex distance.  Thus our chances of finding a
network that does not show the small-world effect are very small for
large~$n$.

\begin{figure}[t]
\begin{center}
\resizebox{8.8cm}{!}{\includegraphics{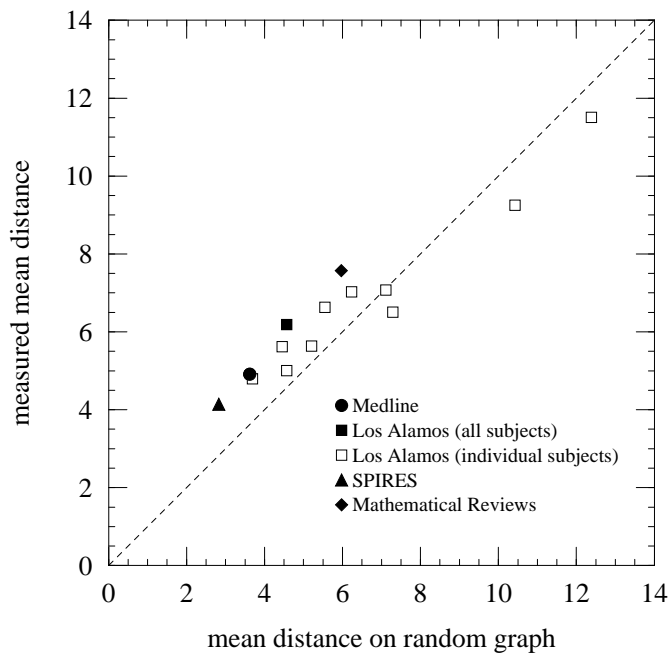}}
\end{center}
\caption{Comparison of mean vertex--vertex distance measured in fourteen
  collaboration networks against our theoretical predictions of the same
  quantities from Eq.~\eref{ell}.  The networks are constructed using
  bibliographic data for papers in biology and medicine (Medline), physics
  (Los Alamos E-print Archive), high-energy physics (SPIRES), and
  mathematics (Mathematical Reviews).  If empirical results and theory
  agreed perfectly, the points would fall on the dotted diagonal line.
  After Newman~\citey{Newman01c}.}
\label{z1z2size}
\end{figure}

As a test of Eq.~\eref{ell}, Fig.~\ref{z1z2size} compares our predictions
of average distance~$\ell$ with direct measurements for fourteen different
scientific collaboration networks, including the biology and mathematics
networks of Table~\ref{clustering}.  In this figure, each network is
represented by a single point, whose position along the horizontal axis
corresponds to the theoretically predicted value of $\ell$ and along the
vertical axis the measured value.  If Eq.~\eref{ell} were exactly correct,
all the points in the figure would fall on the dotted diagonal line.  Since
we know that the equation is only approximate, it comes as no surprise that
the points do not fall perfectly along this line, but the results are
encouraging nonetheless; in most cases the theoretical prediction is close
to the correct result and the overall scaling of $\ell$ with $\log n$ is
clear.  If the theory were equally successful for networks of other types,
it would provide a useful way of estimating average vertex--vertex
separation.  Since $z_1$ and $z_2$ are local quantities that can be
calculated at least approximately from measurements on only a small portion
of the network, it would in many cases be considerably simpler and more
practical to apply Eq.~\eref{ell} than to measure~$\ell$ directly.

Although our random graph model does not allow us to fix the level of
clustering in the network, we can still calculate an average clustering
coefficient for the Bender--Canfield ensemble easily enough.  Consider a
particular vertex~A again.  The $i$th neighbour of~A has $k_i$ edges
emerging from it other than the edge attached to~A, and $k_i$ is
distributed according to the distribution $q_k$, Eq.~\eref{defsqk}.  The
probability that this vertex is connected to another neighbour $j$ is $k_i
k_j/(nz)$, where $k_j$ is also distributed according to~$q_k$, and average
of this probability is precisely the clustering coefficient:
\begin{equation}
C = {\av{k_i k_j}\over nz}
  = {1\over nz}\,\Bigl[\sum_k k q_k\Bigr]^2
  = {z\over n}\,\biggl[{\av{k^2}-\av{k}\over\av{k}^2}\biggr]^2
  = {z\over n}\,\biggl[c_v^2+{z-1\over z}\biggr]^2.
\label{crg}
\end{equation}
The quantity $c_v$ is the so-called coefficient of variation of the degree
distribution---the ratio of the standard deviation to the mean.  Thus the
clustering coefficient for the random graph with a non-Poisson degree
distribution is equal to its value $z/n$ for the Poisson-distributed case,
times a function whose leading term goes as the fourth power of the
coefficient of variation of the degree distribution.  So the clustering
coefficient still vanishes with increasing graph size, but may have a much
larger leading coefficient, since $c_v$ can be quite large, especially for
degree distributions with long tails, such as those seen in
Fig.~\ref{degree}.

Take for example the World-Wide Web.  If one ignores the directed nature of
links on the Web, then the resulting graph is measured to have quite a high
clustering coefficient of $0.11$ \citep{Adamic99}, as shown in
Table~\ref{clustering}.  The Erd\H{o}s--R\'enyi random graph with the same
$n$ and~$z$, by contrast, has a clustering coefficient of only $0.00023$.
However, if we use the degree distribution shown in Fig.~\ref{degree}a to
calculate a mean degree and coefficient of variation for the Web, we get
$z=10.23$ and $c_v=3.685$, which means that $(c_v^2+(z-1)/z)^2=209.7$.
Eq.~\eref{crg} then tells us that the random graph with the correct degree
distribution would actually have a clustering coefficient of
$C=0.00023\times209.7=0.048$.  This is still about a factor of two away
from the correct answer, but a lot closer to the mark than the original
estimate, which was off by a factor of more than~400.  Furthermore, the
degree distribution used in this calculation was truncated at $k=4096$.
(The data were supplied to author in this form.)  Without this truncation,
the coefficient of variation would presumably be larger still.  It seems
possible therefore, that most, if not all, of the clustering seen in the
Web can be accounted for merely as a result of the long-tailed degree
distribution.  Thus the fact that our random graph models do not explicitly
include clustering is not necessarily a problem.

On the other hand, some of the other networks of Table~\ref{clustering} do
show significantly higher clustering than would be predicted by
Eq.~\eref{crg}.  For these, our random graphs will be an imperfect model,
although as we will see they still have much to contribute.  Extension of
our models to include clustering explicitly is discussed in
Section~\ref{ripples}.

\medbreak
It would be possible to continue the analysis of our random graph models
using the simple methods of this section.  However, this leads to a lot of
tedious algebra which can be avoided by introducing an elegant tool, the
probability generating function.

\section{Probability generating functions}
\label{gfs}
In this section we describe the use of probability generating functions to
calculate the properties of random graphs.  Our presentation closely
follows that of Newman~\etal~\citey{NSW01}.

A \defn{probability generating function} is an alternative representation
of a probability distribution.  Take the probability distribution $p_k$
introduced in the previous section, for instance, which is the distribution
of vertex degrees in a graph.  The corresponding generating function is
\begin{equation}
G_0(x) = \sum_{k=0}^\infty p_k x^k.
\label{defsg0}
\end{equation}
It is clear that this function captures all of the information present in
the original distribution~$p_k$, since we can recover $p_k$ from $G_0(x)$
by simple differentiation:
\begin{equation}
p_k = {1\over k!} {\d^k G_0\over\d x^k}\bigg|_{x=0}.
\label{derivatives}
\end{equation}
We say that the function $G_0$ ``generates'' the probability
distribution~$p_k$.

We can also define a generating function for the distribution~$q_k$,
Eq.~\eref{defsqk}, of other edges leaving the vertex we reach by following
an edge in the graph:
\begin{equation}
G_1(x) = \sum_{k=0}^\infty q_k x^k 
       = {\sum_{k=0}^\infty (k+1) p_{k+1}x^k\over\sum_j j p_j}
       = {\sum_{k=0}^\infty k p_k x^{k-1}\over\sum_j j p_j}
       = {G_0'(x)\over z},
\label{defsg1}
\end{equation}
where $G_0'(x)$ denotes the first derivative of $G_0(x)$ with respect to
its argument.  This generating function will be useful to us in following
developments.

\subsection{Properties of generating functions}
\label{properties}
Generating functions have some properties that will be of use in this
paper.  First, if the distribution they generate is properly normalized
then
\begin{equation}
G_0(1) = \sum_k p_k = 1.
\label{normalization}
\end{equation}
Second, the mean of the distribution can be calculated directly by
differentiation:
\begin{equation}
G'_0(1) = \sum_k k p_k = \av{k}.
\label{mean}
\end{equation}
Indeed we can calculate any moment of the distribution by taking a suitable
derivative.  In general,
\begin{equation}
\av{k^n} = \sum_k k^n p_k = 
  \biggl[ \biggl(x {\d\over\d x}\biggr)^{\!n} G_0(x) \biggr]_{x=1}.
\label{moments}
\end{equation}

Third, and most important, if a generating function generates the
probability distribution of some property $k$ of an object, such as the
degree of a vertex, then the sum of that property over $n$ independent such
objects is distributed according to the $n$th power of the generating
function.  Thus the sum of the degrees of $n$ randomly chosen vertices on
our graph has a distribution which is generated by the function
$[G_0(x)]^n$.  To see this, note that the coefficient of $x^m$ in
$[G_0(x)]^n$ has one term of the form $p_{k_1}p_{k_2}\ldots p_{k_n}$ for
every set $\set{k_i}$ of the degrees of the $n$ vertices such that $\sum_i
k_i=m$.  But these terms are precisely the probabilities that the degrees
sum to~$m$ in every possible way, and hence $[G_0(x)]^n$ is the correct
generating function.  This property is the reason why generating functions
are useful in the study of random graphs.  Most of the results of this
paper rely on it.

\subsection{Examples}
\label{examples}
To make these ideas more concrete, let us consider some specific examples
of generating functions.  Suppose for instance that we are interested in
the standard Erd\H{o}s--R\'enyi random graph, with its Poisson degree
distribution.  Substituting Eq.~\eref{poisson} into~\eref{defsg0}, we get
\begin{equation}
G_0(x) = \e^{-z} \sum_{k=0}^\infty {z^k\over k!} x^k
       = \e^{z(x-1)}.
\end{equation}
This is the generating function for the Poisson distribution.  The
generating function $G_1(x)$ for vertices reached by following an edge is
also easily found, from Eq.~\eref{defsg1}:
\begin{equation}
G_1(x) = {G_0'(x)\over z} = \e^{z(x-1)}.
\end{equation}
Thus, for the case of the Poisson distribution we have $G_1(x)=G_0(x)$.
This identity is the reason why the properties of the Erd\H{o}s--R\'enyi
random graph are particularly simple to solve analytically.\footnote{This
  result is also closely connected to our earlier result that the mean
  number of second neighbours of a vertex on an Erd\H{o}s--R\'enyi graph is
  simply the square of the mean number of first neighbours.}

As a second example, consider a graph with an exponential degree
distribution:
\begin{equation}
p_k = (1 - \e^{-1/\kappa}) \e^{-k/\kappa},
\label{expdist}
\end{equation}
where $\kappa$ is a constant.  The generating function for this
distribution is
\begin{equation}
G_0(x) = (1 - \e^{-1/\kappa}) \sum_{k=0}^\infty \e^{-k/\kappa} x^k
       = {1 - \e^{-1/\kappa}\over 1 - x\e^{-1/\kappa}},
\end{equation}
and
\begin{equation}
G_1(x) = \biggl[ {1 - \e^{-1/\kappa}\over 1 - x\e^{-1/\kappa}} \biggr]^2.
\end{equation}

As a third example, consider a graph in which all vertices have degree~0,
1, 2, or~3 with probabilities $p_0\ldots p_3$.  Then the generating
functions take the form of simple polynomials
\begin{eqnarray}
G_0(x) &=& p_3 x^3 + p_2 x^2 + p_1 x + p_0,\\
G_1(x) &=& q_2 x^2 + q_1 x + q_0
        =  {3 p_3 x^2 + 2 p_2 x + p_1\over 3 p_3 + 2 p_2 + p_1}.
\label{g0g1poly}
\end{eqnarray}

\section{Properties of undirected graphs}
\label{undirected}
We now apply our generating functions to the calculation of a variety of
properties of undirected graphs.  In Section~\ref{directed} we extend the
method to directed graphs as well.

\subsection{Distribution of component sizes}
\label{sizes}
The most basic property we will consider is the distribution of the sizes
of connected components of vertices in the graph.  Let us suppose for the
moment that we are below the phase transition, in the regime in which there
is no giant component.  (We will consider the regime above the phase
transition in a moment.)  As discussed in Section~\ref{rg}, the
calculations will depend crucially on the fact that our graphs do not have
significant clustering.  Instead, the clustering coefficient---the
probability that two of your friends are also friends of one another---is
given by Eq.~\eref{crg}, which tends to zero as $n\to\infty$.  The
probability of any two randomly chosen vertices $i$ and $j$ with degrees
$k_i$ and $k_j$ being connected is the same regardless of where the
vertices are.  It is always equal to $k_i k_j/(nz)$, and hence also tends
to zero as $n\to\infty$.  This means that \emph{any finite component of
  connected vertices has no closed loops in it,} and this is the crucial
property that makes exact solutions possible.  In physics jargon, all
finite components are \defn{tree-like}.

\begin{figure}[b]
\begin{center}
\resizebox{8.8cm}{!}{\includegraphics{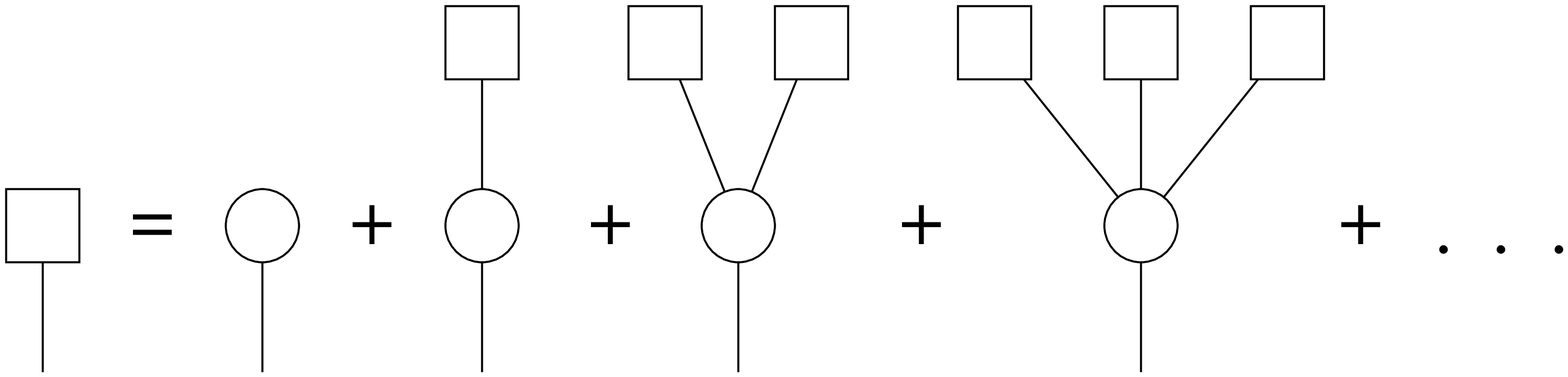}}
\end{center}
\caption{Schematic representation of the possible forms for the connected
  component of vertices reached by following a randomly chosen edge.  The
  total probability of all possible forms (left-hand side) can be
  represented self-consistently as the sum of the probabilities (right-hand
  side) of having only a single vertex (the circle), having a single vertex
  connected to one other component, or two other components, and so forth.
  The entire sum can be expressed in closed form as Eq.~\eref{defsh1}.}
\label{sum}
\end{figure}

Given this, we can calculate the distribution of component sizes below the
transition as follows.  Consider a randomly chosen edge somewhere in our
graph and imagine following that edge to one of its ends and then to every
other vertex reachable from that end.  This set of vertices we refer to as
the cluster at the end of a randomly chosen edge.  Let $H_1(x)$ be the
generating function that generates the distribution of sizes of such
clusters, in terms of numbers of vertices.  Each cluster can take many
different forms, as shown in Fig.~\ref{sum}.  We can follow our randomly
chosen edge and find only a single vertex at its end, with no further edges
emanating from it.  Or we can find a vertex with one or more edges
emanating from it.  Each edge then leads to another complete cluster whose
size is also distributed according to~$H_1(x)$.

The number of edges $k$ emanating from our vertex, other than the one along
which we arrived, is distributed according to the distribution $q_k$ of
Eq.~\eref{defsqk}, and, using the multiplication property of generating
functions from Section~\ref{properties}, the distribution of the sum of the
sizes of the $k$ clusters that they lead to is generated by~$[H_1(x)]^k$.
Thus the total number of vertices reachable by following our randomly
chosen edge is generated by
\begin{equation}
H_1(x) = x \sum_{k=0}^\infty q_k [H_1(x)]^k = x G_1(H_1(x)),
\label{defsh1}
\end{equation}
where the leading factor of $x$ accounts for the one vertex at the end of
our edge, and we have made use of Eq.~\eref{defsg1}.

The quantity we actually want to know is the distribution of the sizes of
the clusters to which a randomly chosen vertex belongs.  The number of
edges emanating from such a vertex is distributed according to the degree
distribution~$p_k$, and each such edge leads to a cluster whose size in
vertices is drawn from the distribution generated by the function $H_1(x)$
above.  Thus the size of the complete component to which a randomly vertex
belongs is generated by
\begin{equation}
H_0(x) = x \sum_{k=0}^\infty p_k [H_1(x)]^k = x G_0(H_1(x)).
\label{defsh0}
\end{equation}
Now we can calculate the complete distribution of component sizes by
solving~\eref{defsh1} self-consistently for $H_1(x)$ and then substituting
the result into~\eref{defsh0}.

Consider for instance the third example from Section~\ref{examples}, of a
graph in which all vertices have degree three or less.  Then
Eq.~\eref{defsh1} implies that $u=H_1(x)$ is a solution of the quadratic
equation
\begin{equation}
q_2 u^2 + \left(q_1-{1\over x}\right) u + q_0 = 0,
\end{equation}
or
\begin{equation}
H_1(x) = {{1\over x} - q_1 
          \pm \sqrt{\bigl(q_1-{1\over x}\bigr)^2-4q_0q_2}\over2q_2}.
\end{equation}
Substituting this into Eq.~\eref{defsh0} and differentiating $m$ times then
gives the probability that a randomly chosen vertex belongs to a component
of exactly $m$ vertices total.

Unfortunately, cases such as this in which we can solve exactly for
$H_0(x)$ and $H_1(x)$ are rare.  More often no closed-form solution is
possible.  (For the simple Poissonian case of the Erd\H{o}s--R\'enyi random
graph, for instance, Eq.~\eref{defsh1} is transcendental and has no
closed-form solution.)  We can still find closed-form expressions for the
generating functions up to any finite order in $x$ however, by iteration
of~\eref{defsh1}.  To see this, suppose that we have an approximate
expression for $H_1(x)$ that is correct up to some finite order~$x^m$, but
possibly incorrect at order $x^{m+1}$ and higher.  If we substitute this
approximate expression into the right-hand side of Eq.~\eref{defsh1}, we
get a new expression for $H_1(x)$ and, because of the leading factor
of~$x$, the only contributions to the coefficient of $x^{m+1}$ in this
expression come from the coefficients of $x^m$ and lower in the old
expression.  Since these lower coefficients were exactly correct, it
immediately follows that the coefficient of $x^{m+1}$ in the new expression
is correct also.  Thus, if we start with the expression $H_1(x)=q_0 x$,
which is correct to order~$x^1$, substitute it into~\eref{defsh1}, and
iterate, then on each iteration we will generate an expression for $H_1(x)$
that is accurate to one order higher.  After $m$ iterations, we will have
an expression in which the coefficients for all orders up to and including
$x^{m+1}$ are exactly correct.

Take for example the Erd\H{o}s--R\'enyi random graph with its Poisson
degree distribution, for which $G_0(x)=G_1(x)=\e^{z(x-1)}$, as shown in
Section~\ref{examples}.  Then, noting that $q_0=\e^{-z}$ for this case, we
find that the first few iterations of Eq.~\eref{defsh1} give
\begin{subequations}
\begin{eqnarray}
zH_1^{(1)}(x) &=& xz\e^{-z} + \O(x^2),\\
zH_1^{(2)}(x) &=& xz\e^{-z} + (xz\e^{-z})^2 + \O(x^3),\\
              &\vdots& \nonumber\\
zH_1^{(5)}(x) &=& xz\e^{-z} + (xz\e^{-z})^2 + \mbox{$\frac32$} (xz\e^{-z})^3
                  + \mbox{$\frac53$} (xz\e^{-z})^4
                  + \mbox{$\frac83$} (xz\e^{-z})^5 + \O(x^6),\nonumber\\
\end{eqnarray}
\end{subequations}
and so forth, from which we conclude that the probabilities $P_s$ of a
randomly chosen site belonging to components of size $s=1,2,3\ldots$ are
\begin{equation}
P_1 = \e^{-z},\quad
P_2 = z\e^{-2z},\quad
P_3 = \mbox{$\frac32$}z^2\e^{-3z},\quad
P_4 = \mbox{$\frac53$}z^3\e^{-4z},\quad
P_5 = \mbox{$\frac83$}z^4\e^{-5z}.
\end{equation}

With a good symbolic manipulation program it is straightforward to
calculate such probabilities to order 100 or so.  If we require
probabilities to higher order it is still possible to use
Eqs.~\eref{defsh1} and~\eref{defsh0} to get answers, by
iterating~\eref{defsh1} numerically from a starting value of $H_1(x)=q_0
x$.  Doing this for a variety of different values of $x$ close to $x=0$, we
can use the results to calculate the derivatives of $H_0(x)$ and so
evaluate the~$P_s$.  Unfortunately, this technique is only usable for the
first few $P_s$, because, as is usually the case with numerical
derivatives, limits on the precision of floating-point numbers result in
large errors at higher orders.  To circumvent this problem we can employ a
technique suggested by Moore and Newman~\citey{MN00b}, and evaluate the
derivatives instead by numerically integrating the Cauchy formula
\begin{equation}
P_s = {1\over s!} {\partial^s H_0\over\partial x^s}\biggr|_{x=0}
    = {1\over2\pi\i} \oint {H_0(\zeta)\>\d\zeta\over\zeta^s},
\end{equation}
where the integral is performed around any contour surrounding the origin
but inside the first pole in~$H_0(\zeta)$.  For the best precision, Moore
and Newman suggest using the largest such contour possible.  In the present
case, where $P_s$ is a properly normalized probability distribution, it is
straightforward to show that $H_0(\zeta)$ must always converge within the
unit circle and hence we recommend using this circle as the contour.  Doing
so appears to give excellent results in practice~\citep{NSW01}, with a
thousand or more derivatives easily calculable in reasonable time.

\subsection{Mean component size}
Although, as we have seen, it is not usually possible to calculate the
probability distribution of component sizes~$P_s$ to all orders in closed
form, we can calculate moments of the distribution, which in many cases is
more useful anyway.  The simplest case is the first moment, the mean
component size.  As we saw in Section~\ref{properties}, the mean of the
distribution generated by a generating function is given by the derivative
of the generating function evaluated at unity (Eq.~\eref{mean}).  Below the
phase transition, the component size distribution is generated by $H_0(x)$,
Eq.~\eref{defsh0}, and hence the mean component size below the transition
is
\begin{equation}
\av{s} = H_0'(1) = \bigl[ G_0(H_1(x)) + x G_0'(H_1(x)) H_1'(x) \bigr]_{x=1}
       = 1 + G_0'(1) H_1'(1),
\label{avs1}
\end{equation}
where we have made use of the fact, Eq.~\eref{normalization}, that properly
normalized generating functions are equal to~1 at $x=1$, so that
$G_0(1)=H_1(1)=1$.  The value of $H_1'(1)$ we can calculate from
Eq.~\eref{defsh1} by differentiating and rearranging to give
\begin{equation}
H_1'(1) = {1\over1-G_1'(1)},
\end{equation}
and substituting into~\eref{avs1} we find
\begin{equation}
\av{s} = 1 + {G_0'(1)\over1-G_1'(1)}.
\label{avs2}
\end{equation}
This expression can also be written in a number of other forms.  For
example, we note that
\begin{eqnarray}
G_0'(1) &=& \sum_k k p_k = \av{k} = z_1,\\
G_1'(1) &=& {\sum_k k(k-1) p_k\over\sum_k k p_k}
         =  {\av{k^2}-\av{k}\over\av{k}} = {z_2\over z_1},
\end{eqnarray}
where we have made use of Eq.~\eref{z2}.  Substituting into~\eref{avs2}
then gives the average component size below the transition as
\begin{equation}
\av{s} = 1 + {z_1^2\over z_1-z_2}.
\end{equation}

This expression has a divergence at $z_1=z_2$, which signifies the
formation of the giant component and gives an alternative and more rigorous
derivation of the position of the critical point to that given in
Section~\ref{rg}.  Using Eq.~\eref{avs2}, we could also write the condition
for the phase transition as $G_1'(1)=1$.

\subsection{Above the phase transition}
The calculations of the previous sections concerned the behaviour of the
graph below the phase transition where there is no giant component in the
graph.  Almost all graphs studied empirically seem to be in the regime
above the transition and do have a giant component.  (This may be a
tautologous statement, since it probably rarely occurs to researchers to
consider a network representation of a set of objects or people so loosely
linked that there is no connection between most pairs.)  Can our generating
function techniques be extended to this regime?  As we now show, they can,
although we will have to use some tricks to make things work.  The problem
is that the giant component is not a component like those we have
considered so far.  Those components had a finite average size, which meant
that in the limit of large graph size they were all tree-like, containing
no closed loops, as discussed in Section~\ref{sizes}.  The giant component,
on the other hand, scales, by definition, as the size of the graph as a
whole, and therefore becomes infinite as $n\to\infty$.  This means that
there will in general be loops in the giant component, which makes all the
arguments of the previous sections break down.  This problem can be fixed
however by the following simple ploy.  Above the transition, we define
$H_0(x)$ and $H_1(x)$ to be the generating functions for the distributions
of component sizes \emph{excluding} the giant component.  The non-giant
components are still tree-like even above the transition, so
Eqs.~\eref{defsh1} and~\eref{defsh0} are correct for this definition.  The
only difference is that now $H_0(1)$ is no longer equal to~1 (and neither
is $H_1(1)$).  Instead,
\begin{equation}
H_0(1) = \sum_s P_s = \mbox{fraction of vertices not in giant component},
\end{equation}
which follows because the sum over $s$ is now over only the non-giant
components, so the probabilities $P_s$ no longer add up to~1.  This result
is very useful; it allows us to calculate the size~$S$ of the giant
component above the transition as a fraction of the total graph size, since
$S=1-H_0(1)$.  From Eqs.~\eref{defsh1} and~\eref{defsh0}, we can see that
$S$ must be the solution of the equations
\begin{equation}
S = 1 - G_0(v),\qquad v = G_1(v),
\label{defsu}
\end{equation}
where $v\equiv H_1(1)$.  As with the calculation of the component size
distribution in Section~\ref{sizes}, these equations are not normally
solvable in closed form, but a solution can be found to arbitrary numerical
accuracy by iteration starting from a suitable initial value of~$v$, such
as $v=0$.

We can also calculate the average sizes of the non-giant components in the
standard way by differentiating Eq.~\eref{defsh0}.  We must be careful
however, for a couple of reasons.  First, we can no longer assume that
$H_0(1)=H_1(1)=1$ as is the case below the transition.  Second, since the
distribution $P_s$ is not normalized to~1, we have to perform the
normalization ourselves.  The correct expression for the average component
size is
\begin{eqnarray}
\av{s} &=& {H_0'(1)\over H_0(1)}
        =  {1\over H_0(1)} \biggl[ G_0(H_1(1)) +
           {G_0'(H_1(1)) G_1(H_1(1))\over1 - G_1'(H_1(1))} \biggr]\nonumber\\
       &=& 1 + {zv^2\over [1-S][1-G_1'(v)]},
\end{eqnarray}
where~$v$ and~$S$ are found from Eq.~\eref{defsu}.  It is straightforward
to verify that this becomes equal to Eq.~\eref{avs2} when we are below the
transition and $S=0$, $v=1$.

\begin{figure}
\begin{center}
\resizebox{8.8cm}{!}{\includegraphics{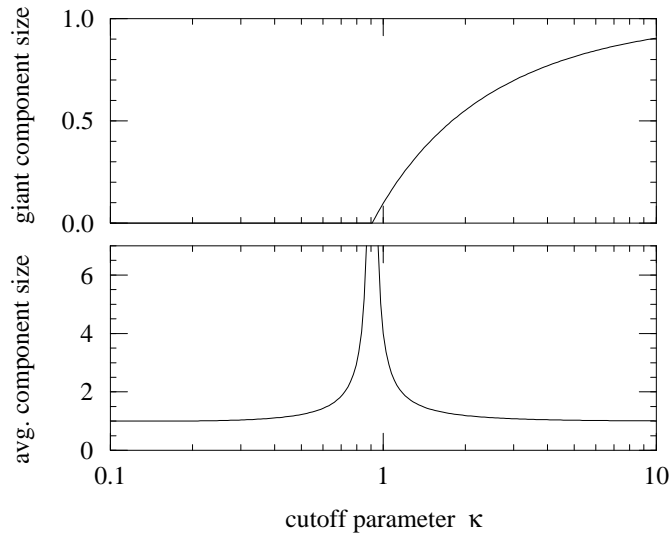}}
\end{center}
\caption{Behaviour of a random graph with an exponential degree
  distribution of the form of Eq.~\eref{expdist}.  Top: fraction of the
  graph occupied by the giant component.  Bottom: average component size.
  Note that the horizontal axis is logarithmic.}
\label{gcavs}
\end{figure}

As an example of these results, we show in Fig.~\ref{gcavs} the size of the
giant component and the average (non-giant) component size for graphs with
an exponential degree distribution of the form of Eq.~\eref{expdist}, as a
function of the exponential constant~$\kappa$.  As the figure shows, there
is a divergence in the average component size at the phase transition, with
the giant component becoming non-zero smoothly above the transition.  Those
accustomed to the physics of continuous phase transitions will find this
behaviour familiar; the size of the giant component acts as an order
parameter here, as it did in the Erd\H{o}s--R\'enyi random graph in the
introduction to this paper, and the average component size behaves like a
susceptibility.  Indeed one can define and calculate critical exponents for
the transition using this analogy, and as with the Erd\H{o}s--R\'enyi
model, their values put us in the same universality class as the mean-field
(i.e.,~infinite dimension) percolation transition \citep{NSW01}.  The phase
transition in Fig.~\ref{gcavs} takes place just a little below $\kappa=1$
when $G_1'(1)=1$, which gives a critical value of
$\kappa_c=(\log3)^{-1}=0.910\ldots$

\section{Properties of directed graphs}
\label{directed}
Some of the graphs discussed in the introduction to this paper are directed
graphs.  That is, the edges in the network have a direction to them.
Examples are the World-Wide Web, in which hyperlinks from one page to
another point in only one direction, and food webs, in which predator--prey
interactions are asymmetric and can be thought of as pointing from predator
to prey.  Other recently studied examples of directed networks include
telephone call graphs \citep{ABW98,Hayes00,ACL00}, citation networks
\citep{Redner98,Vazquez01}, and email networks \citep{EMB02}.

Directed networks are more complex than their undirected counterparts.  For
a start, each vertex in an directed network has two degrees, an
\defn{in-degree}, which is the number of edges that point into the vertex,
and an \defn{out-degree}, which is the number pointing out.  There are
also, correspondingly, two degree distributions.  In fact, to be completely
general, we must allow for a \defn{joint degree distribution} of in- and
out-degree: we define $p_{jk}$ to be the probability that a randomly chosen
vertex simultaneously has in-degree~$j$ and out-degree~$k$.  Defining a
joint distribution like this allows for the possibility that the in- and
out-degrees may be correlated.  For example in a graph where every vertex
had precisely the same in- and out-degree, $p_{jk}$~would be non-zero if
and only if $j=k$.

The component structure of a directed graph is more complex than that of an
undirected graph also, because a directed path may exist through the
network from vertex~A to vertex~B, but that does not guarantee that one
exists from B to~A.  As a result, any vertex~A belongs to components of
four different types:
\begin{enumerate}
\item The \defn{in-component} is the set of vertices from which A can be
  reached.
\item The \defn{out-component} is the set of vertices which can be reached
  from~A.
\item The \defn{strongly connected component} is the set of vertices from
  which vertex~A can be reached \emph{and} which can be reached from~A.
\item The \defn{weakly connected component} is the set of vertices that can
  be reached from~A ignoring the directed nature of the edges altogether.
\end{enumerate}
The weakly connected component is just the normal component to which A
belongs if one treats the graph as undirected.  Clearly the details of
weakly connected components can be worked out using the formalism of
Section~\ref{undirected}, so we will ignore this case.  For vertex~A to
belong to a strongly connected component of size greater than one, there
must be at least one other vertex that can both be reached from~A and from
which A can be reached.  This however implies that there is a closed loop
of directed edges in the graph, something which, as we saw in
Section~\ref{sizes}, does not happen in the limit of large graph size.  So
we ignore this case also.  The two remaining cases, the in- and
out-components, we consider in more detail in the following sections.

\subsection{Generating functions}
Because the degree distribution $p_{jk}$ for a directed graph is a function
of two variables, the corresponding generating function is also:
\begin{equation}
\cG(x,y) = \sum_{j,k=0}^\infty p_{jk} x^j y^k.
\label{jointgen}
\end{equation}
This function satisfies the normalization condition $\cG(1,1)=1$, and the
means of the in- and out-degree distributions are given by its first
derivatives with respect to $x$ and~$y$.  However, there is \emph{only one}
mean degree $z$ for a directed graph, since every edge must start and end
at a site.  This means that the total and hence also the average numbers of
in-going and out-going edges are the same.  This gives rise to a constraint
on the generating function of the form
\begin{equation}
{\partial\cG\over\partial x}\bigg|_{x,y=1} = z =
{\partial\cG\over\partial y}\bigg|_{x,y=1},
\end{equation}
and there is a corresponding constraint on the probability distribution
$p_{jk}$ itself, which can be written
\begin{equation}
\sum_{jk} (j-k) p_{jk} = 0.
\label{inoutfix}
\end{equation}

From $\cG(x,y)$, we can now define single-argument generating functions
$G_0$ and $G_1$ for the number of out-going edges leaving a randomly chosen
vertex, and the number leaving the vertex reached by following a randomly
chosen edge.  These play a similar role to the functions of the same name
in Section~\ref{undirected}.  We can also define generating functions $F_0$
and $F_1$ for the number of edges arriving at a vertex.  These functions
are given by
\begin{eqnarray}
\label{defsf0f1}
F_0(x) &=& \cG(x,1),\qquad
F_1(x) = %{\partial_y \cG(x,1)\over\partial_y \cG(1,1)} =
{1\over z}\, {\partial \cG\over\partial y}\bigg|_{y=1},\\
\label{defsg0g1}
G_0(y) &=& \cG(1,y),\hspace{7.5mm}
G_1(y) = %{\partial_x \cG(1,y)\over\partial_x \cG(1,1)} =
{1\over z}\, {\partial \cG\over\partial x}\bigg|_{x=1}.
\end{eqnarray}
Once we have these functions, many results follow as before.

\subsection{Results}
The probability distribution of the numbers of vertices reachable from a
randomly chosen vertex in a directed graph---i.e.,~of the sizes of the
out-components---is generated by the function $H_0(y) = y G_0(H_1(y))$,
where $H_1(y)$ is a solution of $H_1(y) = y G_1(H_1(y))$, just as before.
(A similar and obvious pair of equations governs the sizes of the
in-components.)  The average out-component size for the case where there is
no giant component is then given by Eq.~\eref{avs2}, and thus the point at
which a giant component first appears is given once more by $G_1'(1) = 1$.
Substituting Eq.~\eref{defsg0g1} into this expression gives the explicit
condition
\begin{equation}
\sum_{jk} (2jk-j-k) p_{jk} = 0
\label{directedpt}
\end{equation}
for the first appearance of the giant component.  This expression is the
equivalent for the directed graph of Eq.~\eref{mrcondition}.  It is also
possible, and equally valid, to define the position at which the giant
component appears by $F_1'(1)=1$, which provides an alternative derivation
for Eq.~\eref{directedpt}.

But this raises an interesting issue.  Which giant component are we talking
about?  Just as with the small components, there are four types of giant
component, the giant in- and out-components, and the giant weakly and
strongly connected components.  Furthermore, while the giant weakly
connected component is as before trivial, the giant strongly connected
component does not normally vanish as the other strongly connected
components do.  There is no reason why a giant component should contain no
loops, and therefore no reason why we should not have a non-zero giant
strongly connected component.

The condition for the position of the phase transition given above is
derived from the point at which the mean size of the out-component
reachable from a vertex diverges, and thus this is the position at which
the giant \emph{in}-component forms (since above this point an extensive
number of vertices can be reached starting from one vertex, and hence that
vertex must belong to the giant in-component).  Furthermore, as we have
seen, we get the same condition if we ask where the mean in-component size
diverges, i.e.,~where the giant \emph{out}-component forms, and so we
conclude that both giant in- and out-components appear at the same time, at
the point given by Eq.~\eref{directedpt}.

The sizes of these two giant components can also be calculated with only a
little extra effort.  As before, we can generalize the functions $H_0(y)$
and $H_1(y)$ to the regime above the transition by defining them to be the
generating functions for the non-giant out-components in this regime.  In
that case, $H_0(1)$ is equal to the fraction of all vertices that have a
finite out-component.  But any vertex~A that has only a finite
out-component cannot, by definition, belong to the giant
\emph{in-}component, i.e.,~there definitely do not exist an extensive
number of vertices that can be reached from~A.  Thus the size of the giant
in-component is simply $S_{\rm in}=1-H_0(1)$, which can be calculated as
before from Eq.~\eref{defsu}.  Similarly the size of the giant
out-component can be calculated from~\eref{defsu} with $G_0\to F_0$ and
$G_1\to F_1$.

To calculate the size of the giant strongly connected component, we observe
the following \citep{DMS01a}.  If at least one of a vertex's outgoing edges
leads to anywhere in the giant \emph{in}-component, then one can reach the
giant strongly connected component from that vertex.  Conversely, if at
least one of a vertex's incoming edges leads from anywhere in the giant
\emph{out}-component, then the vertex can be reached from the strongly
connected component.  If and only if both of these conditions are satisfied
simultaneously, then the vertex belongs to the giant strongly connected
component itself.

Consider then the outgoing edges.  The function $H_1(x)$ gives the
probability distribution of the sizes of finite out-components reached by
following a randomly chosen edge.  This implies that $H_1(1)$ is the total
probability that an edge leads to a finite out-component (i.e.,~\emph{not}
to the giant in-component) and as before (Eq.~\eref{defsu}) $H_1(1)$ is the
fixed point of $G_1(x)$, which we denote by~$v$.  For a vertex with $k$
outgoing edges, $v^k$~is then the probability that all of them lead to
finite components and $1-v^k$ is the probability that at least one edge
leads to the giant in-component.  Similarly the probability that at least
one incoming edge leads from the giant out-component is $1-u^j$, where $u$
is the fixed point of $F_1(x)$ and $j$ is the in-degree of the vertex.
Thus the probability that a vertex with in- and out-degrees $j$ and $k$ is
in the giant strongly connected component is $(1-u^j)(1-v^k)$, and the
average of this probability over all vertices, which is also the fractional
size of the giant strongly connected component,~is
\begin{eqnarray}
S_s &=& \sum_{jk} p_{jk} (1-u^j)(1-v^k)
     =  \sum_{jk} p_{jk} (1-u^j-v^k+u^j v^k)\nonumber\\
    &=& 1 - \cG(u,1) - \cG(1,v) + \cG(u,v),
\end{eqnarray}
where $u$ and $v$ are solutions of
\begin{equation}
u = F_1(u),\qquad v = G_1(v),
\end{equation}
and we have made use of the definition, Eq.~\eref{jointgen}, of $\cG(x,y)$.
Noting that $u=v=1$ below the transition at which the giant in- and
out-components appear, and that $\cG(1,1)=1$, we see that the giant
strongly connected component also first appears at the transition point
given by Eq.~\eref{directedpt}.  Thus there are in general two phase
transitions in a directed graph: the one at which the giant weakly
connected component appears, and the one at which the other three giant
components all appear.

Applying the theory of directed random graphs to real directed networks has
proved difficult so far, because experimenters rarely measure the joint in-
and out-degree distribution $p_{jk}$ that is needed to perform the
calculations described above.  A few results can be calculated without the
joint distribution---see Newman~\etal~\citey{NSW01}, for instance.  By and
large, however, the theory presented in this section is still awaiting
empirical tests.

\section{Networks with clustering}
\label{ripples}
Far fewer analytical results exist for networks that incorporate clustering
than for the non-clustered networks of the previous sections.  A first
attempt at extending random graph models to incorporate clustering has been
made by the present author, who studied the correction to the quantity
$z_2$---the average number of next-nearest neighbours of a vertex---in
graphs with a non-zero clustering coefficient~$C$ \citep{Newman02a}.

Consider a vertex~A, with its first and second neighbours in the network
arrayed around it in two concentric rings.  In a normal random graph, a
neighbour of~A that has degree $m$ contributes $m-1$ vertices to the ring
of second neighbours of~A, as discussed in Section~\ref{rg}.  That is, all
of the second neighbours of~A are independent; each of them is a new vertex
never before seen.  This is the reasoning that led to our earlier
expression, Eq.~\eref{z2}: $z_2=\av{k^2}-\av{k}$.  In a clustered network
however, the picture is different.  In a clustered network, many of the
neighbours of A's neighbour are neighbours of~A themselves.  This is the
meaning of clustering: your friend's friend is also your friend.  In fact,
by definition, an average fraction $C$ of the $m-1$ neighbours are
themselves neighbours of the central vertex~A and hence should not be
counted as second neighbours.  Correspondingly, this reduces our estimate
of $z_2$ by a factor of $1-C$ to give $z_2=(1-C)(\av{k^2}-\av{k})$.

But this is not all.  There is another effect we need to take into account
if we are to estimate $z_2$ correctly.  It is also possible that we are
overcounting the second neighbours of~A because some of them are neighbours
of more than one of the first neighbours.  In other words, you may know two
people who have another friend in common, whom you personally don't know.
Such connections create ``squares'' in the network, whose density can be
quantified by the so-called \defn{mutuality}~$M$:
\begin{equation}
M = {\mbox{mean number of vertices two steps away from a vertex}\over
     \mbox{mean number of paths of length two to those vertices}}.
\label{defsm}
\end{equation}
In words, $M$ measures the average number of paths of length two leading to
a vertex's second neighbour.  As a result of the mutuality effect, our
current estimate of $z_2$ will be too great by a factor of~$1/M$, and hence
a better estimate is
\begin{equation}
z_2 = M(1-C)(\av{k^2}-\av{k}).
\label{betterz2}
\end{equation}

\begin{figure}
\begin{center}
\resizebox{6.6cm}{!}{\includegraphics{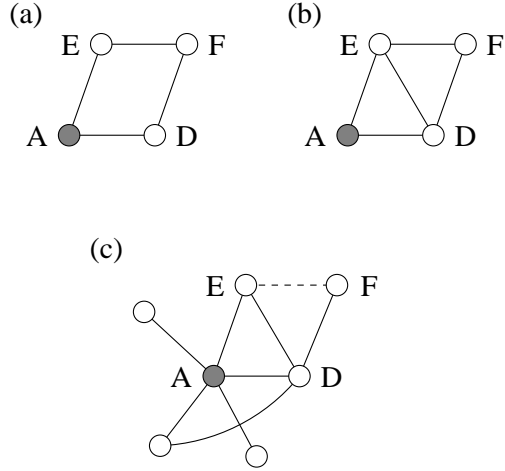}}
\end{center}
\caption{(a)~An example of a vertex~(F) that is two steps away from the
  center vertex~(A, shaded), but is connected to two of A's neighbours
  (D~and~E).  F~should only be counted once as a second neighbour of~A, not
  twice.  (b)~A similar situation in which D~and~E are also neighbours of
  one another.  (c)~The probability of situation (b) can be calculated by
  considering this situation.  Since D is friends with both E and~F, the
  probability that E and F also know one another (dotted line), thereby
  completing the quadrilateral in~(b), is by definition equal to the
  clustering coefficient.}
\label{arrangement}
\end{figure}

But now we have a problem.  Calculating the mutuality $M$ using
Eq.~\eref{defsm} requires that we know the mean number of individuals two
steps away from the central vertex~A.  But this mean number is precisely
the quantity $z_2$ that our calculation is supposed to estimate in the
first place.  There is a partial solution to this problem.  Consider the
two configurations depicted in Fig.~\ref{arrangement}, parts~(a) and~(b).
In~(a) our vertex~A has two neighbours D and~E, both of whom are connected
to~F, although F is not itself an neighbour of~A.  The same is true in~(b),
but now D and~E are friends of one another also.  Empirically, it appears
that in many networks situation~(a) is quite uncommon, while situation~(b)
is much more common.  And we can estimate the frequency of occurrence of
(b) from a knowledge of the clustering coefficient.

Consider Fig.~\ref{arrangement}c.  The central vertex~A shares an edge
with~D, which shares an edge with~F.  How many other paths of length two
are there from~A to~F?  Well, if A has $k_1$ neighbours, then by the
definition of the clustering coefficient, D~will be connected to $C(k_1-1)$
of them on average.  The edge between vertices~D and~E in the figure is an
example of one such.  But now~D is connected to both E and~F, and hence,
using the definition of the clustering coefficient again, E~and~F will
themselves be connected (dotted line) with probability equal to the
clustering coefficient~$C$.  Thus there will on average be $C^2(k_1-1)$
other paths of length~2 to~F, or $1+C^2(k_1-1)$ paths in total, counting
the one that runs through~D.  This is the average factor by which we will
overcount the number of second neighbours of~A because of the mutuality
effect.  As shown by Newman~\citey{Newman02a}, the mutuality coefficient is
then given by
\begin{equation}
M = {\av{k/[1+C^2(k-1)]}\over\av{k}}.
\label{mutuality}
\end{equation}
Substituting this into Eq.~\eref{betterz2} then gives us an estimate
of~$z_2$.

In essence what Eq.~\eref{mutuality} does is estimate the value of $M$ in a
network in which triangles of ties are common, but squares that are not
composed of adjacent triangles are assumed to occur with frequency no
greater than one would expect in a purely random network.  It is only an
approximate expression, since this assumption will usually not be obeyed
perfectly.  Nonetheless, it appears to give good results.  The author
applied Eqs.~\eref{betterz2} and~\eref{mutuality} to estimation of $z_2$
for the two coauthorship networks of Fig.~\ref{degree}c, and found that
they gave results accurate to within 10\% in both cases.

This calculation is certainly only a first step.  Ideally we would like to
be able to calculate numbers of vertices at any distance from a randomly
chosen central vertex in the presence of clustering, and to do it exactly
rather than just approximately.  If this were possible, then, as in
Section~\ref{rg}, one could use the ratio of the numbers of vertices at
different distances to derive a condition for the position of the phase
transition at which a giant component forms on a clustered graph.  At
present it is not clear if such a calculation is possible.

\section{Models defined on random graphs}
\label{models}
In addition to providing an analytic framework for calculating topological
properties of networks, such as typical path lengths or distributions of
cluster sizes, random graphs form a useful substrate for studying the
behaviour of phenomena that take place on networks.  Analytic work in this
area is in its infancy; here we describe two examples of recent work on
models that use ideas drawn from percolation theory.

\subsection{Network resilience}
\label{resilience}
As emphasized by Albert and co-workers, the highly skewed degree
distributions of Fig.~\ref{degree} have substantial implications for the
robustness of networks to the removal of vertices \citep{AJB00}.  Because
most of the vertices in a network with such a degree distribution typically
have low degree, the random removal of vertices from the network has little
effect on the connectivity of the remaining vertices, i.e.,~on the
existence of paths between pairs of vertices, a crucial property of
networks such as the Internet, for which functionality relies on
connectivity.\footnote{A few recent papers in the physics literature have
  used the word ``connectivity'' to mean the same thing as ``degree'',
  i.e.,~number of edges attaching to a vertex.  In this paper however the
  word has its standard graph theoretical meaning of existence of
  connecting paths between pairs of vertices.}  In particular, removal of
vertices with degree zero or one will never have any effect on the
connectivity of the remaining vertices.  (Vertices of degree zero are not
connected to anyone else anyway, and vertices of degree one do not lie on
any path between another pair of vertices.)

Conversely, however, the specific removal of the vertices in the network
with the highest degree frequently has a devastating effect.  These
vertices lie on many of the paths between pairs of other vertices and their
removal can destroy the connectivity of the network in short order.  This
was first demonstrated numerically by Albert~\etal~\citey{AJB00} and
independently by Broder~\etal~\citey{Broder00} using data for subsets of
the World-Wide Web.  More recently however it has been demonstrated
analytically also, for random graphs with arbitrary degree distributions,
by Callaway~\etal~\citey{CNSW00} and by Cohen~\etal~\citey{CEBH01}.  Here
we follow the derivation of Callaway~\etal, which closely mirrors some of
the earlier mathematical developments of this paper.

Consider a simple model defined on a network in which each vertex is either
``present'' or ``absent''.  Absent vertices are vertices that have either
been removed, or more realistically are present but non-functional, such as
Internet routers that have failed or Web sites whose host computer has gone
down.  We define a probability $b_k$ of being present which is some
arbitrary function of the degree~$k$ of a vertex, and then define the
generating function
\begin{equation}
F_0(x)=\sum_{k=0}^{\infty} p_k b_k x^k,
\label{defsf0}
\end{equation}
whose coefficients are the probabilities that a vertex has degree~$k$ and
is present.  Note that this generating function is not equal to~1 at $x=1$;
instead it is equal to the fraction of all vertices that are present.  By
analogy with Eq.~\eref{defsg1} we also define
\begin{equation}
F_1(x)=\frac{\sum_k kp_k b_k x^{k-1}}{\sum_k kp_k} = \frac{F_0'(x)}{z}.
\end{equation}
Then the distributions of the sizes of connected clusters of present
vertices reachable from a randomly chosen vertex or edge are generated
respectively by
\begin{equation} 
H_0(x) = 1 - F_0(1) + x F_0(H_1(x)),\qquad
H_1(x) = 1 - F_1(1) + x F_1(H_1(x)),
\label{fullh0h1}
\end{equation}
which are logical equivalents of Eqs.~\eref{defsh1} and~\eref{defsh0}.

Take for instance the case of random failure of vertices.  In this case,
the probability $b_k$ of a vertex being present is independent of the
degree~$k$ and just equal to a constant~$b$, which means that
\begin{equation}
H_0(x) = 1 - b + b x G_0(H_1(x)),\qquad
H_1(x) = 1 - b + b x G_1(H_1(x)),
\label{siteperc}
\end{equation}
where $G_0(x)$ and $G_1(x)$ are the standard generating functions for
vertex degree, Eqs.~\eref{defsg0} and~\eref{defsg1}.  This implies that the
mean size of a cluster of connected and present vertices is
\begin{equation}
\av{s} = H_0'(1) = b + b F_0'(1) H_1'(1)
       = b\left[1+\frac{bG_0'(1)}{1-bG_1'(1)}\right],
\end{equation}
and the model has a phase transition at the critical value of~$b$
\begin{equation}
b_c=\frac{1}{G_1'(1)}.
\end{equation}
If a fraction $b<b_c$ of the vertices are present in the network, then
there will be no giant component.  This is the point at which the network
ceases to be functional in terms of connectivity.  When there is no giant
component, connecting paths exist only within small isolated groups of
vertices, but no long-range connectivity exists.  For a communication
network such as the Internet, this would be fatal.  As we would expect from
the arguments above however, $b_c$~is usually a very small number for
networks with skewed degree distributions.  For example, if a network has a
pure power-law degree distribution with exponent~$\alpha$, as both the
Internet and the World-Wide Web appear to do (see Fig.~\ref{degree}a
and~\ref{degree}b), then
\begin{equation}
b_c = {\zeta(\alpha-1)\over\zeta(\alpha-2)-\zeta(\alpha-1)},
\end{equation}
where $\zeta(x)$ is the Riemann $\zeta$-function.  This expression is
formally zero for all $\alpha\le3$.  Since none of the distributions in
Fig.~\ref{degree} have an exponent greater than~3, it follows that, at
least to the extent that these graphs can be modelled as random graphs,
none of them has a phase transition at all.  No matter how many vertices
fail in these networks, as long as the failing vertices are selected at
random without regard for degree, there will always be a giant component in
the network and an extensive fraction of the vertices will be connected to
one another.  In this sense, networks with power-law distributed degrees
are highly robust, as the numerical experiments of
Albert~\etal~\citey{AJB00} and Broder~\etal~\citey{Broder00} also found.

But now consider the case in which the vertices are removed in decreasing
order of their degrees, starting with the highest degree vertex.
Mathematically we can represent this by setting
\begin{equation}
b_k=\theta(k_{\rm max}-k),
\end{equation}
where $\theta(x)$ is the Heaviside step function
\begin{equation}
\theta(x) = \biggl\lbrace \begin{array}{ll}
            0             & \mbox{for $x<0$}\\
            1 \qquad\null & \mbox{for $x\ge0$.}
            \end{array}
\end{equation}
This is equivalent to setting the upper limit of the sum in
Eq.~\eref{defsf0} to $k_{\rm max}$.

For this case we need to use the full definition of $H_0(x)$ and $H_1(x)$,
Eq.~\eref{fullh0h1}, which gives the position of the phase transition as
the point at which $F_1'(1)=1$, or
\begin{equation}
{\sum_{k=1}^\infty k(k-1) p_k b_k\over\sum_{k=1}^\infty k p_k} = 1.
\end{equation}
Taking the example of our power-law degree distribution again, $p_k\propto
k^{-\alpha}$, this then implies that the phase transition occurs at a value
$k_c$ of $k_{\rm max}$ satisfying
\begin{equation}
H_{k_c}^{(\alpha-2)}-H_{k_c}^{(\alpha-1)} = \zeta(\alpha-1),
\label{defskc}
\end{equation}
where $H_n^{(r)}$ is the $n$th harmonic number of order~$r$:
\begin{equation}
H_n^{(r)} = \sum_{k=1}^n {1\over k^r}.
\end{equation}

\begin{figure}
\begin{center}
\resizebox{8.8cm}{!}{\includegraphics{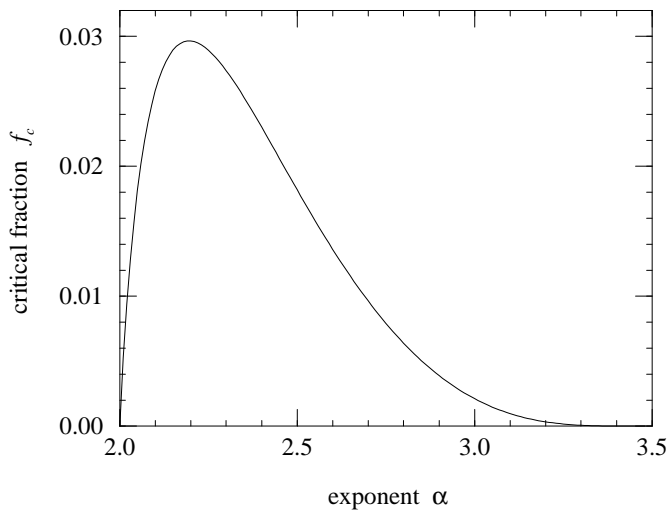}}
\end{center}
\caption{The critical fraction, Eq.~\eref{defsfc}, of highest degree
  vertices that must be removed in order to destroy the giant component in
  a graph with a power-law degree distribution having exponent~$\alpha$.}
\label{figfc}
\end{figure}

This solution is not in a very useful form however.  What we really want to
know is what fraction $f_c$ of the vertices have been removed when we reach
the transition.  This fraction is given by
\begin{equation}
f_c = 1 - {H_{k_c}^{(\alpha)}\over\zeta(\alpha)}.
\label{defsfc}
\end{equation}
Although we cannot eliminate $k_c$ from~\eref{defskc} and~\eref{defsfc} to
get $f_c$ in closed form, we can solve Eq.~\eref{defskc} numerically for
$k_c$ and substitute into~\eref{defsfc}.  The result is shown as a function
of $\alpha$ in Fig.~\ref{figfc}.  As the figure shows, one need only remove
a very small fraction of the high-degree vertices to destroy the giant
component in a power-law graph, always less than~3\%, with the most robust
graphs being those around $\alpha=2.2$, interestingly quite close to the
exponent seen in a number of real-world networks (Fig.~\eref{degree}).
Below $\alpha=2$, there is no real solution for $f_c$: power-law
distributions with $\alpha<2$ have no finite mean anyway and therefore make
little sense physically.  And $f_c=0$ for all values $\alpha>3.4788\ldots$,
where the latter figure is the solution of
$\zeta(\alpha-2)=2\zeta(\alpha-1)$, because the underlying network itself
has no giant component for such values of~$\alpha$ \citep{ACL00}.

Overall, therefore, our results agree with the findings of the previous
numerical studies that graphs with skewed degree distributions, such as
power laws, can be highly robust to the random removal of vertices, but
extremely fragile to the specific removal of their highest-degree vertices.

\subsection{Epidemiology}
An important application of the theory of networks is in epidemiology, the
study of the spread of disease.  Diseases are communicated from one host to
another by physical contact, and the pattern of who has contact with whom
forms a \defn{contact network} whose structure has implications for the
shape of epidemics.  In particular, the small-world effect discussed in
Section~\ref{rg} means that diseases will spread through a community much
faster than one might otherwise imagine.

In the standard mathematical treatments of diseases, researchers use the
so-called \defn{fully mixed approximation}, in which it is assumed that
every individual has equal chance of contact with every other.  This is an
unrealistic assumption, but it has proven popular because it allows one to
write differential equations for the time evolution of the disease that can
be solved or numerically integrated with relative ease.  More realistic
treatments have also been given in which populations are divided into
groups according to age or other characteristics.  These models are still
fully mixed within each group however.  To go beyond these approximations,
we need to incorporate a full network structure into the model, and the
random graphs of this paper and the generating function methods we have
developed to handle them provide a good basis for doing this.

In this section we show that the most fundamental standard model of disease
propagation, the SIR model, and a large set of its generalized forms, can
be solved on random graphs by mapping them onto percolation problems.
These solutions provide exact criteria for deciding when an epidemic will
occur, how many people will be affected, and how the network structure or
the transmission properties of the disease could be modified in order to
prevent the epidemic.

\subsection{The SIR model}
First formulated (though never published) by Lowell Reed and Wade Hampton
Frost in the 1920s, the SIR model \citep{Bailey75,AM91,Hethcote00} is a
model of disease propagation in which members of a population are divided
into three classes: susceptible~(S), meaning they are free of the disease
but can catch it; infective~(I), meaning they have the disease and can pass
it on to others;\footnote{In common parlance, the word ``infectious'' is
  more often used, but in the epidemiological literature ``infective'' is
  the accepted term.} and removed~(R), meaning they have recovered from the
disease or died, and can no longer pass it on.  There is a fixed
probability per unit time that an infective individual will pass the
disease to a susceptible individual with whom they have contact, rendering
that individual infective.  Individuals who contract the disease remain
infective for a certain time period before recovering (or dying) and
thereby losing their infectivity.

As first pointed out by Grassberger~\citey{Grassberger83}, the SIR model on
a network can be simply mapped to a bond percolation process.  Consider an
outbreak on a network that starts with a single individual and spreads to
encompass some subset of the network.  The vertices of the network
represent potential hosts and the edges represent pairs of hosts who have
contact with one another.  If we imagine occupying or colouring in all the
edges that result in transmission of the disease during the current
outbreak, then the set of vertices representing the hosts infected in this
outbreak form a connected percolation cluster of occupied edges.
Furthermore, it is easy to convince oneself that each edge is occupied with
independent probability.  If we denote by $\tau$ the time for which an
infected host remains infective and by $r$ the probability per unit time
that that host will infect one of its neighbours in the network, then the
total probability of infection is
\begin{equation}
T = 1 - \lim_{\delta t\to0} (1-r\,\delta t)^{\tau/\delta t}
  = 1 - \e^{-r\tau}.
\label{defstrans1}
\end{equation}
This quantity we call the \defn{transmissibility}, and it is the
probability that any edge on the network is occupied.  The size
distribution of outbreaks of the disease is then given by the size
distribution of percolation clusters on the network when edges are occupied
with this probability.  When the mean cluster size diverges, we get
outbreaks that occupy a finite fraction of the entire network,
i.e.,~epidemics; the percolation threshold corresponds to what an
epidemiologist would call the \defn{epidemic threshold} for the disease.
Above this threshold, there exists a giant component for the percolation
problem, whose size corresponds to the size of the epidemic.  Thus, if we
can solve bond percolation on our random graphs, we can also solve the SIR
model.

In fact, we can also solve a generalized form of the SIR in which both
$\tau$ and $r$ are allowed to vary across the network.  If $\tau$ and $r$
instead of being constant are picked at random for each vertex or edge from
some distributions $P(\tau)$ and $P(r)$, then the probability of
percolation along any edge is simply the average of Eq.~\eref{defstrans1}
over these two distributions \citep{WSS01,Warren02}:
\begin{equation}
T = 1 - \int P(r) P(\tau)\, \e^{-r\tau}\>\d r\>\d\tau.
\label{defstrans2}
\end{equation}

\subsection{Solution of the SIR model}
The bond percolation problem on a random graph can be solved by techniques
very similar to those of Section~\ref{resilience} \citep{CNSW00,Newman02b}.
The equivalent of Eq.~\eref{siteperc} for bond percolation with bond
occupation probability~$T$ is
\begin{equation}
H_0(x) = x G_0(H_1(x)),\qquad H_1(x) = 1 - T + T x G_1(H_1(x)),
\end{equation}
which gives an average outbreak size below the epidemic threshold of
\begin{equation}
\av{s} = H_0'(1) = 1 + {T G_0'(1)\over 1 - T G_1'(1)}.
\label{outbreak}
\end{equation}
The threshold itself then falls at the point where $TG_1'(1)=1$, giving a
critical transmissibility of
\begin{equation}
T_c = {1\over G_1'(1)} = {\av{k}\over\av{k^2}-\av{k}} = {z_1\over z_2},
\label{epidemic}
\end{equation}
where we have used Eq.~\eref{z2}.  The size~$S$ of the epidemic above the
epidemic transition can be calculated by finding the solution of
\begin{equation}
S = 1 - G_0(v),\qquad v = 1 - T + T G_1(v),
\end{equation}
which will normally have to be solved numerically, since closed form
solutions are rare.  It is also interesting to ask what the probability is
that an outbreak starting with a single carrier will become an epidemic.
This is precisely equal to the probability that the carrier belongs to the
giant percolating cluster, which is also just equal to~$S$.  The
probability that a given infection event (i.e.,~transmission along a given
edge) will give rise to an epidemic is $v\equiv H_1'(1)$.

Newman and co-workers have given a variety of further generalizations of
these solutions to networks with structure of various kinds, models in
which the probabilities of transmission between pairs of hosts are
correlated in various ways, and models incorporating vaccination, either
random or targeted, which is represented as a site percolation process
\citep{ANMS02,Newman02b}.  To give one example, consider the network by
which a sexually transmitted disease is communicated, which is also the
network of sexual partnerships between individuals.  In a recent study of
2810 respondents, Liljeros~\etal~\citey{Liljeros01} recorded the numbers of
sexual partners of men and women over the course of a year.  From their
data it appears that the distributions of these numbers follow a power law
similar to those of the distributions in Fig.~\ref{degree}, with exponents
$\alpha$ that fall in the range $3.1$ to~$3.3$.  If we assume that the
disease of interest is transmitted primarily by contacts between men and
women (true only for some diseases), then to a good approximation the
network of contacts is bipartite, having two separate sets of vertices
representing men and women and edges representing contacts running only
between vertices of unlike kinds.  We define two pairs of generating
functions for males and females:
\begin{eqnarray}
F_0(x) &=& \sum_j p_j x^j,\qquad
F_1(x) = \frac{1}{\mu} \sum_j j p_j x^{j-1},\\
G_0(x) &=& \sum_k q_k x^k,\hspace{6.8mm}
G_1(x) = \frac{1}{\nu} \sum_k k q_k x^{k-1},
\end{eqnarray}
where $p_j$ and $q_k$ are the two degree distributions and $\mu$ and $\nu$
are their means.  We can then develop expressions similar to
Eqs.~\eref{outbreak} and~\eref{epidemic} for an epidemic on this new
network.  We find, for instance, that the epidemic transition takes place
at the point where $T_{mf} T_{fm} = 1/[F_1'(1)G_1'(1)]$ where $T_{mf}$ and
$T_{fm}$ are the transmissibilities for male-to-female and female-to-male
infection respectively.

One important result that follows immediately is that if the degree
distributions are truly power-law in form, then there exists an epidemic
transition only for a small range of values of the exponent $\alpha$ of the
power law.  Let us assume, as appears to be the case \citep{Liljeros01},
that the exponents are roughly equal for men and women:
$\alpha_m=\alpha_f=\alpha$.  Then if $\alpha\le3$, we find that $T_{mf}
T_{fm} = 0$, which is only possible if at least one of the
transmissibilities $T_{mf}$ and $T_{fm}$ is zero.  As long as both are
positive, we will always be in the epidemic regime, and this would clearly
be bad news.  No amount of precautionary measures to reduce the probability
of transmission would ever eradicate the disease.  (Similar results have
been seen in other types of models also \citep{PV01a,LM01}.)  Conversely,
if $\alpha>\alpha_c$, where $\alpha_c=3.4788\ldots$ is the solution of
$\zeta(\alpha-2)=2\zeta(\alpha-1)$, we find that $T_{mf} T_{fm}>1$, which
is not possible.  (This latter result arises because networks with
$\alpha>\alpha_c$ have no giant component at all, as mentioned in
Section~\ref{resilience} \citep{ACL00}.)  In this regime then, no epidemic
can ever occur, which would be good news.  Only in the small intermediate
region $3<\alpha<3.4788\ldots$ does the model possess an epidemic
transition.  Interestingly, the real-world network measured by
Liljeros~\etal~\citey{Liljeros01} appears to fall precisely in this region,
with $\alpha\simeq3.2$.  If true, this would be both good and bad news.  On
the bad side, it means that epidemics can occur.  But on the good side, it
means that that it is in theory possible to prevent an epidemic by reducing
the probability of transmission, which is precisely what most health
education campaigns attempt to do.  The predicted critical value of the
transmissibility is $\zeta(\alpha-1)/[\zeta(\alpha-2)-\zeta(\alpha-1)]$,
which gives $T_c=0.363\ldots$ for $\alpha=3.2$.  Epidemic behaviour would
cease were it possible to arrange that $T_{mf}T_{fm}<T_c^2$.

\section{Summary}
In this paper we have given an introduction to the use of random graphs as
models of real-world networks.  We have shown (Section~\ref{rg}) how the
much studied random graph model of Erd\H{o}s and R\'enyi can be generalized
to the case of arbitrary degree distributions, allowing us to mimic the
highly skewed degree distributions seen in many networks.  The resulting
models can be solved exactly using generating function methods in the case
where there is no clustering (Sections~\ref{gfs} and~\ref{undirected}).  If
clustering is introduced, then solutions become significantly harder, and
only a few approximate analytic results are known (Section~\ref{ripples}).
We have also given solutions for the properties of directed random graphs
(Section~\ref{directed}), in which each edge has a direction that it points
in.  Directed graphs are useful as models of the World-Wide Web and food
webs, amongst other things.  In the last part of this paper
(Section~\ref{models}) we have given two examples of the use of random
graphs as a substrate for models of dynamical processes taking place on
networks, the first being a model of network robustness under failure of
vertices (e.g.,~failure of routers on the Internet), and the second being a
model of the spread of disease across the network of physical contacts
between disease hosts.  Both of these models can be mapped onto percolation
problems of one kind of another, which can then be solved exactly, again
using generating function methods.

There are many conceivable extensions of the theory presented in this
paper.  In particular, there is room for many more and diverse models of
processes taking place on networks.  It would also be of great interest if
it proved possible to extend the results of Section~\ref{ripples} to obtain
exact or approximate estimates of the global properties of networks with
non-zero clustering.

\section*{Acknowledgements}
The author thanks Duncan Callaway, Peter Dodds, Michelle Girvan, Andr\'e
Krzywicki, Len Sander, Steve Strogatz and Duncan Watts for useful and
entertaining conversations.  Thanks are also due to Jerry Davis, Paul
Ginsparg, Jerry Grossman, Oleg Khovayko, David Lipman, Heath O'Connell,
Grigoriy Starchenko, Geoff West and Janet Wiener for providing data used in
some of the examples.  The original research described in this paper was
supported in part by the US National Science Foundation under grant number
DMS--0109086.

\vspace{0.5cm}
\begin{center}
\rule{5cm}{0.5pt}
\end{center}

\end{document}